\documentclass[
aps,%
final,%
oneside,%
onecolumn,%
nobibnotes,%
nofootinbib,%
superscriptaddress,%
showpacs,%
centertags]%
{revtex4}
\usepackage{mathrsfs}
\usepackage{amssymb,amsmath,amscd, amsthm,stmaryrd,verbatim, mathrsfs}

\newtheorem{Th}{Theorem}
\newtheorem{Def}{Definition}
\newtheorem{Lem}{Lemma}

\newtheorem{Prop}{Proposition}
\newtheorem{rmk}{Remark}

\newcommand{\bb}{\mathbb}
\newcommand{\ms}{\mathscr}
\newcommand{\mr}{\mathrm}
\newcommand{\frk}{\mathfrak}

\usepackage{hyperref}
\begin{document}

\title{Generalized MICZ-Kepler Problems and Unitary Highest Weight Modules}

\author{\firstname{Guowu}~\surname{Meng}}
\email{mameng@ust.hk} \affiliation{Department of Mathematics, Hong
Kong Univ. of Sci. and Tech., Clear Water Bay, Kowloon, Hong Kong}
\author{\firstname{Ruibin}~\surname{Zhang}}
\email{rzhang@mail.usyd.edu.au} \affiliation{School of Mathematics
and Statistics, University of Sydney Sydney, NSW 2006, Australia}

\begin{abstract}
For each integer $n\ge 1$, we demonstrate that a
$(2n+1)$-dimensional generalized MICZ-Kepler problem has an
$\mr{Spin}(2, 2n+2)$ dynamical symmetry which extends the manifest
$\mr{Spin}(2n+1)$ symmetry. The Hilbert space of bound states is
shown to form a unitary highest weight $\mr{Spin}(2, 2n+2)$-module
which occurs at the first reduction point in the
Enright-Howe-Wallach classification diagram for the unitary highest
weight modules. As a byproduct, we get a simple geometric
realization for such a unitary highest weight $\mr{Spin}(2,
2n+2)$-module.
\end{abstract}
 \maketitle

\tableofcontents

\section {Introduction}
The Kepler problem is a physics problem in dimension three about two
bodies which attract each other by a force proportional to the
inverse square of their distance.  As is well known, its exact
solution in classical mechanics gives a very satisfactory
explanation of the Kepler's laws of planetary motion, and its exact
solution in quantum mechanics gives an equally satisfactory
explanation of the spectral lines for the hydrogen atom. The
MICZ-Kepler problems, discovered in the late 60s by McIntosh and
Cisneros \cite{MC70} and independently by Zwanziger \cite{Z68},  are
natural cousins of the Kepler problem. Roughly speaking, a
MICZ-Kepler problem is the Kepler problem for which the nucleus of a
hypothetic hydrogen atom also carries a magnetic charge.

In the early 90s, Iwai \cite{Iwai90} obtained non-abelian analogues
of the MICZ-Kepler problems in dimension five; more recently, the
first author constructed and solved \cite{meng05} analogues of the
MICZ-Kepler problems in all dimensions bigger than or equal to three
which extends the aforementioned work of McIntosh and Cisneros,
Zwanziger, and Iwai. We shall refer to the MICZ-Kepler problems and
their higher dimensional analogues as the \emph{generalized
MICZ-Kepler problems}.

Recall that the MICZ-Kepler problems all have a large dynamical
symmetry group--$\mr{Spin}(2, 4)$--as shown by Barut and Bornzin
\cite{Barut71}. These authors also used the symmetry to provide an
elegant solution for the problems in \cite{Barut71}. Similar results
were also established in dimension five by Pletyukhov and Tolkachev
\cite{Pletyukhov99} for the generalized MICZ-Kepler problems of
Iwai. The purpose of the present paper is to investigate the
dynamical symmetry and explore its representation theory for the
generalized MICZ-Kepler problems in all odd dimensions.

We shall show that for each positive integer $n$, a
$(2n+1)$-dimensional generalized MICZ-Kepler problem always has an
$\mr{Spin}(2, 2n+2)$\footnote{It is a double cover of $\mr{SO}_0(2,
2n+2)$ --- the identity component of $\mr{SO}(2, 2n+2)$. By
definition, it is characterized by the homomorphism
$\pi_1(\mr{SO}_0(2, 2n+2))=\bb Z\oplus \bb Z_2\to \bb Z_2$ which
maps $(a, b)$ to $\bar a+b$. Here $\bar a$ is the congruence class
of $a$ modulo $2$.} dynamical symmetry, i.e., \emph{its Hilbert
space of bound states forms an irreducible unitary  module for
$\mr{Spin}(2, 2n+2)$}. In fact, we shall show that the Hilbert space
of bound states forms a unitary highest weight module for
$\mr{Spin}(2, 2n+2)$; more precisely, we shall establish the
following result\footnote{The result for an even dimensional
generalized MICZ-Kepler problem shall be presented in Ref.
\cite{MZ07b}}:

\begin{Th}\label{main} Assume $n\ge 1$ is an integer and $\mu$ is an
half integer. Let ${\ms H}(\mu)$ be the Hilbert space of bound
states for the $(2n+1)$-dimensional generalized MICZ-Kepler problem
with magnetic charge $\mu$, and $l_\mu=l+|\mu|+n-1$ for any integer
$l\ge 0$.

1) There is a natural unitary action of $\mr{Spin}(2, 2n+2)$ on $
{\ms H}(\mu)$ which extends the manifest unitary action of
$\mr{Spin}(2n+1)$. In fact, ${\ms H}(\mu)$ is the unitary highest
weight module of $\mr{Spin}(2, 2n+2)$ with highest weight
$\left(-(n+|\mu|),|\mu|, \cdots, |\mu|, \mu\right)$; consequently,
it occurs at the first reduction point of the Enright-Howe-Wallach
classification diagram\footnote{Page 101, Ref. \cite{EHW82}. See
also Refs. \cite{Jakobsen81a,Jakobsen81b}.} for the unitary highest
weight modules, so it is a non-discrete series representation.

2) As a representation of subgroup $\mr{Spin}(2, 1)\times_{\bb Z_2}
\mr{Spin}(2n+1)$,
\begin{eqnarray}
{\ms H}(\mu) = \hat \bigoplus_{l=0}^\infty \left({\cal
D}^-_{2l_\mu+2}\otimes D_l\right)
\end{eqnarray} where $D_l$ is the irreducible module of $\mr{Spin}(2n+1)$
with highest weight $(l+|\mu|, |\mu|, \cdots, |\mu|)$ and ${\cal
D}^-_{2l_\mu+2}$ is the anti-holomorphic discrete series
representation of $\mr{Spin}(2, 1)$ with highest weight $-l_\mu-1$.

3) As a representation of the maximal compact subgroup
$\mr{Spin}(2)\times_{\bb Z_2} \mr{Spin}(2n+2)$,
\begin{eqnarray}
{\ms H}(\mu) = \hat \bigoplus_{l=0}^\infty \left(D(-l_\mu-1)\otimes
D^l\right)
\end{eqnarray} where $D^l$ is the irreducible module of $\mr{Spin}(2n+2)$
with highest weight $(l+|\mu|, |\mu|, \cdots, |\mu|, \mu)$ and
$D(-l_\mu-1)$ is the irreducible module of $\mr{Spin}(2)$ with
weight $-l_\mu-1$.

\end{Th}
Readers who wish to have a quick geometric description of the
aforementioned unitary highest weight module of $\mr{Spin}(2, 2n+2)$
may consult the appendix. Readers who wish to know more details
about the classification \cite{Jakobsen81a, Jakobsen81b, EHW82} of
unitary highest weight modules may start with a fairly readable
account from Ref. \cite{EHW82}. Note that there is no general
classification result for the family of unitary modules of real
non-compact simple Lie groups, and the subfamily of unitary
\emph{highest weight} modules is special enough so that such a nice
classification result can possibly exist. The first reduction point
picked up by the ``Nature" from the Enright-Howe-Wallach
classification diagram is even more special because it belongs to an
even more special subfamily called Wallach set.

\vskip 10pt In section \ref{GMIC}, we give a quick review of the
generalized MICZ-Kepler problems in odd dimensions. For the
computational purpose in the subsequent section, we quickly review
the gauge potential\footnote{It is $\sqrt {-1}$ times the local {\em
connection one-form}. } for the background gauge field (i.e.,
\emph{connection}) under a particular local gauge (i.e., {\em bundle
trivialization}), and then quote from Ref. \cite{meng05} some key
identities satisfied by the gauge potential. In section \ref{HDS},
we introduce the dynamical symmetry operators and show that they
satisfy the commutation relations for the generators\footnote{Here
we adopt the practice in physics: the Lie algebra generators act as
hermitian operators in all unitary representations.} of
$\frk{so}_0(2, 2n+2)$. We also show that these dynamical symmetry
operators satisfy a set of quadratic relations \footnote{This set of
quadratic relations will be shown \cite{meng07} to algebraically
characterize the unitary highest weight ${\mr{Spin}}(2,
2n+2)$-modules stated in Theorem \ref{main} above.}. In section
\ref{Rep}, we start with a preliminary discussion of the
representation problem and point out the need of ``twisting". Then
we gave a review of the (bound) energy eigenspaces (i.e.,
\emph{eigenspaces of the harmiltonian} viewed as a hermitian
operator on the physical Hilbert space) and finally introduce the
notion of ``twisted'' energy eigenspaces which is soon shown to be
the space of $L^2$-sections of a canonical hermitian bundle. In the
last section, we solve the representation problem by proving two
propositions from which Theorem \ref{main} follows quickly. In the
appendix, each of the unitary highest weight representation of
$\mr{Spin}(2, 2n+2)$ encountered here is geometrically realized as
the space of all $L^2$ sections of a canonical hermitian bundle. Via
communications with Profs. R. Howe and N. Wallach, we learned that
these representations can be imbedded into the kernel of certain
canonical differential operators, see Refs. \cite{Tan-Howe, KO03}
for the case $\mu=0$ and Ref. \cite{EW97} for the general case.
Prof. Feher informed us of Ref. \cite{Feher 86} in which a related
interesting model with a conjectured dynamical $O(2,4)$ symmetry is
investigated.

\section{Review of generalized MICZ-Kepler problems}\label{GMIC}
From the physics point of view, a MICZ-Kepler problem is a
generalization of the Kepler problem by adding a suitable background
magnetic field, while at the same time making an appropriate
adjustment to the scalar Coulomb potential so that the problem is
still integrable. The configuration space is the punctured 3D
Euclidean space, and the background magnetic field is a Dirac
monopole. To be more precise, the (dimensionless) hamiltonian of a
MICZ-Kepler problem with magnetic charge $\mu $ is
\begin{eqnarray}H
 = -\frac{1}{2}{\Delta}_{\mathcal A} + \frac{\mu^2}{2r^2}
 - \frac{1}{r}\;.
\end{eqnarray}
Here $\Delta_{\mathcal A}$ is the Laplace operator twisted by the
gauge potential $\mathcal A$ of a Dirac monopole under a particular
gauge, and $\mu$ is the magnetic charge of the Dirac monopole, which
must be a half integer.

To extend the MICZ-Kepler problems beyond dimension three, one needs
a suitable generalization of the Dirac monopoles. Fortunately this
problem was solved in Refs. \cite{meng04, Cotaescu05, meng05}. We
review the work here.

\subsection{Generalized MICZ-Kepler problems}

Let $D\ge 3$ be an integer, $\bb R^{D}_*$ be the punctured
$D$-space, i.e., $\bb R^D$ with the origin removed. Let $ds^2$ be
the cylindrical metric on $\bb R^D_*$. Then $(\bb R_*^D, ds^2)$ is
the product of the straight line $\bb R$ with the round sphere ${\mr
S}^{D-1}$. Since we are interested in the odd dimensional
generalized MICZ-Kepler problems only in this paper, we assume $D$
is odd.

Let $\mathcal S_\pm$ be the positive/negative spinor bundle of $(\bb
R^D_*, ds^2)$, then $\mathcal S_\pm$ correspond to the fundamental
spin representations ${\bf s}_\pm$ of $\frk{so}_0(D-1)$ (the Lie
algebra of $\mr{SO}(D-1)$). Note that each of the above spinor
bundles is endowed with a natural $\mr{SO}(D)$ invariant connection
--- the Levi-Civita spin connection of $(\bb R^D_*, ds^2)$. As a
result, the Young product of $I$ copies of these bundles, denoted by
${\mathcal S}_+^I$, ${\mathcal S}_-^I$ respectively, are also
equipped with natural $\mr{SO}(D)$ invariant connections.

When $\mu$ is a positive half integer, we write ${\mathcal
S}_+^{2\mu}$ as ${\mathcal S}^{2\mu}$, and ${\mathcal S}_-^{2\mu}$
as ${\mathcal S}^{-2\mu}$. We also adopt this convention for
$\mu=0$: to denote by ${\mathcal S}^{0}$ the product complex line
bundle with the product connection. Note that ${\mathcal S}^{2\mu}$
with $\mu$ being a half integer is our analogue of the Dirac
monopole with magnetic charge $\mu$, and the corresponding
representation of $\frk{so}_0(D-1)$ will be denoted by ${\bf
s}^{2\mu}$.

\begin{Def} Let $n\ge 1$ be an integer, $\mu$ a half integer.
The $(2n+1)$-dimensional generalized MICZ-Kepler problem with
magnetic charge $\mu$ is defined to be the quantum mechanical system
on $\bb R^{2n+1}_*$ for which the wave-functions are sections of
${\mathcal S}^{2\mu}$, and the hamiltonian is
\begin{eqnarray}
H
 = -\frac{1}{2}\Delta_\mu + \frac{(n-1)|\mu|+\mu^2}{2r^2}
 - \frac{1}{r}
\end{eqnarray}
where $\Delta_\mu$ is the Laplace operator twisted by ${\mathcal
S}^{2\mu}$.
\end{Def}

Upon choosing a local gauge, the background gauge field (i.e., the
natural connection on ${\mathcal S}^{2\mu}$) can be represented by a
gauge potential ${\mathcal A}_\alpha$ in an explicit form; then
$\Delta_\mu$ can be represented explicitly by
$\sum_\alpha(\partial_\alpha +i {\mathcal A}_\alpha)^2$. Since the
gauge potential is of crucial importance, we review some of its
properties in the next subsection.

\subsection {Basic identities for the gauge potential}\label{GDM}

We write $\vec r = (x_1, x_2, \ldots, x_{D-1}, x_D)$ for a point in
$\bb R^{D}$ and $r$ for the length of $\vec r$. The small Greek
letters $\mu$, $\nu$, etc run from $1$ to $D$ and the lower case
Latin letters $a$, $b$ etc run from $1$ to $D-1$. We use the
Einstein convention that repeated indices are always summed over.

Under a suitable choice of local gauge on ${\bb R}^{D}$ with the
negative $D$-th axis removed, the gauge field can be represented by
the following gauge potential:
\begin{eqnarray}\label{mnple}
{\mathcal A}_D=0,\hskip 20 pt {\mathcal A}_b=-{1\over
r(r+x_D)}x_a\gamma_{ab}
\end{eqnarray}
where $\gamma_{ab}={i\over 4}[\gamma_a,\gamma_b]$ with $\gamma_a$
being the ``gamma matrix" for physicists.  Note that $\gamma_a=ie_a$
with $e_a$ being the element in the Clifford algebra that
corresponds to the $a$-th standard coordinate vector of $\bb
R^{D-1}$.

The field strength of $\mathcal A_\alpha$ is then given by
\begin{eqnarray}F_{Db}& = & {1\over r^3}x_a\gamma_{ab},\cr
F_{ab}& = & -{2\gamma_{ab}\over r(r+x_D)}+  {1\over
r^2(r+x_D)^2}\cdot\cr & &\left((2+{x_D\over r})x_c(x_a
\gamma_{cb}-x_b \gamma_{ca}) +ix_d x_c[\gamma_{d a},\gamma_{c
b}]\right)\end{eqnarray}

Here are some identities from Ref. \cite{meng05} that our later
computations will crucially depend on:
\begin{Lem}\label{lemma}
Let $\mathcal A_\alpha$ be the gauge potential defined by
equation(\ref{mnple}) and let $F_{\alpha\beta}$ be its field
strength.

1) The following identities are valid in any representation of
$\frk{so}_0(D-1)$:
\begin{eqnarray}\label{Id}
F_{\mu\nu}F^{\mu\nu}=\frac{2}{r^4}c_2,\quad
{[\nabla_\kappa,F_{\mu\nu}]}={1\over r^2}\left( x_\mu
F_{\nu\kappa}+x_\nu F_{\kappa \mu}-2x_\kappa F_{\mu\nu} \right), \cr
x_\mu {\mathcal A}_\mu=0,\hskip 20pt x_\mu F_{\mu\nu}=0, \hskip 20pt
[\nabla_\mu, F_{\mu\nu}]=0, \cr r^2[F_{\mu\nu},
F_{\alpha\beta}]+iF_{\mu\beta}\delta_{\alpha\nu}-
iF_{\nu\beta}\delta_{\alpha\mu}+iF_{\alpha\mu}\delta_{\beta\nu}-iF_{\alpha\nu}\delta_{\beta\mu}\cr
={i\over r^2}\left(x_\mu x_\alpha F_{\beta\nu}+x_\mu x_\beta
F_{\nu\alpha}-x_\nu x_\alpha F_{\beta\mu}-x_\nu x_\beta
F_{\mu\alpha} \right),
\end{eqnarray}
where $\nabla_\alpha=\partial_\alpha+i\mathcal A_\alpha$, and
$c_2=c_2[\frk{so}_0(D-1)]={1\over 2}\gamma_{ab}\gamma_{ab}$ is the
(quadratic) Casimir operator of $\frk{so}_0(D-1)$.

2) When $D=2n+1$, $\mu$ is a half integer, the following identity
\begin{eqnarray}\label{keyID}
r^2F_{\lambda\alpha}F_{\lambda\beta}={c_2\over n}\left({1\over
r^2}\delta_{\alpha\beta}-{x_\alpha x_\beta\over
r^4}\right)+i(n-1)F_{\alpha\beta}
\end{eqnarray} holds in the
irreducible representation ${\bf s}^{2\mu}$ of $\frk{so}_0(2n)$ with
highest weight $(|\mu|, \cdots, |\mu|, \mu)$.
\end{Lem}
Note that ${c_2\over n}=\mu^2+(n-1)|\mu|$ in the irreducible
representation ${\bf s}^{2\mu}$. Remark that ${\mathcal
A}_r={\mathcal A}_\theta=0$, where ${\mathcal A}_r$ and ${\mathcal
A}_\theta$ are the $r$ and $\theta$ components of $\mathcal A$ in
the polar coordinate system $(r, \theta, \theta_1,
\cdots,\theta_{D-3},\phi)$ for $\bb R^D_*$ with $\theta$ being the
angle between $\vec r$ and the positive $D$-th axis.

\section {The dynamical symmetry}\label{HDS}
For the remainder of this paper, we only consider a fixed
$(2n+1)$-dimensional generalized MICZ-Kepler problem with magnetic
charge $\mu$. Recall that the configuration space is $\bb R^D_*$
where $D=2n+1$. For our computational purposes, it suffices to work
on ${\bb R}^D$ with the negative $D$-axis removed. Introduce the
notations $\pi_\alpha:=-i\nabla_\alpha$, $c:=\mu^2+(n-1)|\mu|$. Then
$[\pi_\alpha, \pi_\beta]=-iF_{\alpha\beta}$.

Following Barut and Bornzin \cite{Barut71}, we let
\begin{eqnarray}\label{def1}\left\{\begin{array}{l}
\vec \Gamma  := r\vec \pi, \hskip 10pt X :=r\pi^2+{c\over r}, \hskip
10pt
 Y := r,\cr
 { J}_{\alpha\beta}  := i[\Gamma_\alpha, \Gamma_\beta],\hskip
 10pt
 \vec Z :=
i[\vec \Gamma, X],\hskip 10pt\vec W :=  i[\vec \Gamma, Y]=\vec r;
\end{array}\right.
\end{eqnarray} and
\begin{eqnarray}\label{def2}\left\{\begin{array}{l}
\Gamma_{D+1} := {1\over 2}\left(X-Y \right),  \hskip 30pt
 \Gamma_{-1} := {1\over 2}\left(X+Y \right),
 \\
 \vec A := {1\over 2}\left(\vec Z-\vec W \right),  \hskip 10pt
 \vec M := {1\over 2}\left(\vec Z+\vec W \right),\hskip 10pt
 T :=i[\Gamma_{D+1}, \Gamma_{-1}].\end{array}\right.
\end{eqnarray}
Some relatively straightforward but lengthy computations yield
\begin{eqnarray}\left\{\begin{array}{rcl}
{ J}_{\alpha\beta}  &= & x_\alpha\pi_\beta
-x_\beta\pi_\alpha+r^2F_{\alpha\beta},\cr A_\alpha &= &{1\over
2}x_\alpha\pi^2 - \pi_\alpha(\vec r\cdot \vec \pi)+r^2F_{\alpha
\beta}\pi_\beta-{c\over 2r^2}x_\alpha +{i\over
2}(D-3)\pi_\alpha-{1\over 2}x_\alpha,\cr M_\alpha &= &{1\over
2}x_\alpha\pi^2 - \pi_\alpha(\vec r\cdot \vec \pi)+r^2F_{\alpha
\beta}\pi_\beta-{c\over 2r^2}x_\alpha +{i\over
2}(D-3)\pi_\alpha+{1\over 2}x_\alpha,\cr T  &= & \vec r\cdot \vec
\pi-i{D-1\over 2},\cr \Gamma_\alpha &=& r\pi_\alpha,\cr \Gamma_{-1}
&= & {1\over 2}\left(r\pi^2+r+{c\over r} \right),\cr \Gamma_{D+1} &=
& {1\over 2}\left(r\pi^2-r+{c\over r} \right).\end{array}\right.
\end{eqnarray}

Let the capital Latin letters $A$, $B$ run from $-1$ to $D+1$.
Introduce $J_{AB}$ as follows:
\begin{eqnarray}\label{defofJ}
J_{AB}=\left\{\begin{array}{ll}  J_{\mu\nu} & \hbox{if $A=\mu$,
$B=\nu$}\cr  A_\mu & \hbox{if $A=\mu$, $B=D+1$}\cr M_\mu& \hbox{if
$A=\mu$, $B=-1$}\cr \Gamma_\mu & \hbox{if $A=\mu$, $B=0$}\cr T &
\hbox{if $A=D+1$, $B=-1$}\cr \Gamma_{D+1} & \hbox{if $A=D+1$,
$B=0$}\cr \Gamma_{-1} & \hbox{if $A=-1$, $B=0$}\cr -J_{BA} &
\hbox{if $A>B$}\cr 0 & \hbox{if $A=B$}.\cr
\end{array}\right.
\end{eqnarray}

\begin{Th}\label{keyr} Let $C^\infty({\mathcal S}^{2\mu})$ be the space of smooth sections of ${\mathcal S}^{2\mu}$. Let $J_{AB}$
be defined by (\ref{defofJ}).

1) As operators on $C^\infty({\mathcal S}^{2\mu})$, $J_{AB}$'s
satisfy the following commutation relations:
\begin{eqnarray}\label{cmtr} [J_{AB},
J_{A'B'}]=-i\eta_{AA'}J_{BB'}-i\eta_{BB'}J_{AA'}+i\eta_{AB'}J_{BA'}+i\eta_{BA'}J_{AB'}
\end{eqnarray}
where the indefinite metric tensor $\eta$ is ${\mr
{diag}}\{++-\cdots-\}$ relative to the following order: $-1$, $0$,
$1$, \ldots, $2n+2$ for the indices.

2) As operators on $C^\infty({\mathcal S}^{2\mu})$,
\begin{eqnarray} \{J_{AB}, {J^A}_C\}:=J_{AB}{J^A}_C+{J^A}_CJ_{AB}=-2a\eta_{BC}
\end{eqnarray}where $a=n-c$.
\end{Th}
The proof of this theorem is purely algebraic and computational, but
quite long. It will be carried out in the next two subsections.
\subsection{Proof of part 1)}
By exploiting the symmetry properties of both sides of Eq.
(\ref{cmtr}), we can see that it suffices to verify the commutation
relations in the cases where $(A, B)\neq (A', B')$, $A<B$, $A'<B'$
and $B'\le B$. The proof crucially depends on Lemma \ref{lemma}.

The following lemma is quite useful.
\begin{Lem}\label{lemma2}
\begin{eqnarray}\left\{\begin{array}{rcl}
[J_{\alpha\beta}, r]& = & [J_{\alpha\beta}, {1\over r}]=0,\cr
[J_{\alpha\beta}, x_\nu]& = &
-i(x_\alpha\delta_{\beta\nu}-x_\beta\delta_{\alpha\nu}),\cr
[J_{\alpha\beta}, \pi_\nu]& = &
-i(\pi_\alpha\delta_{\beta\nu}-\pi_\beta\delta_{\alpha\nu}),\cr
[J_{\alpha\beta}, F_{\alpha'\beta'}] &= &
i\delta_{\alpha\alpha'}F_{\beta\beta'}+i\delta_{\beta\beta'}F_{\alpha\alpha'}
-i\delta_{\alpha\beta'}F_{\beta\alpha'}-i\delta_{\beta\alpha'}F_{\alpha\beta'}.\end{array}\right.
\end{eqnarray}
\end{Lem}
\begin{proof}
\begin{eqnarray}
[J_{\alpha\beta}, r]& = &
[x_\alpha\pi_\beta-x_\beta\pi_\alpha+r^2F_{\alpha\beta},r]=[x_\alpha\pi_\beta-x_\beta\pi_\alpha,r]\cr
&= & -i(x_\alpha {x_\beta\over r}-x_\beta{x_\alpha\over
r})=0.\nonumber
\end{eqnarray}
\begin{eqnarray}
[J_{\alpha\beta}, {1\over r}]& = &
[x_\alpha\pi_\beta-x_\beta\pi_\alpha+r^2F_{\alpha\beta},{1\over
r}]=[x_\alpha\pi_\beta-x_\beta\pi_\alpha,{1\over r}]\cr &= &
+i(x_\alpha {x_\beta\over r^3}-x_\beta{x_\alpha\over
r^3})=0.\nonumber
\end{eqnarray}
\begin{eqnarray}
[J_{\alpha\beta}, x_\nu]& = &
[x_\alpha\pi_\beta-x_\beta\pi_\alpha+r^2F_{\alpha\beta},x_\nu]=[x_\alpha\pi_\beta-x_\beta\pi_\alpha,x_\nu]\cr
&= &
-i(x_\alpha\delta_{\beta\nu}-x_\beta\delta_{\alpha\nu}).\nonumber
\end{eqnarray}
\begin{eqnarray}
[J_{\alpha\beta}, \pi_\nu]& = &
[x_\alpha\pi_\beta-x_\beta\pi_\alpha+r^2F_{\alpha\beta},\pi_\nu]\cr
&= &
-i(\pi_\alpha\delta_{\beta\nu}-\pi_\beta\delta_{\alpha\nu})-ix_\alpha
F_{\beta\nu}+ix_\beta F_{\alpha\nu}\cr & & +2ix_\nu
F_{\alpha\beta}+ir^2[\nabla_\nu, F_{\alpha\beta}]\cr &=&
-i(\pi_\alpha\delta_{\beta\nu}-\pi_\beta\delta_{\alpha\nu}).\nonumber
\end{eqnarray}
\begin{eqnarray}
[J_{\alpha\beta},F_{\alpha'\beta'}] & = &
[x_\alpha\pi_\beta-x_\beta\pi_\alpha+r^2F_{\alpha\beta},
F_{\alpha'\beta'}]\cr &=& x_\alpha[\pi_\beta,
F_{\alpha'\beta'}]-x_\beta
[\pi_\alpha,F_{\alpha'\beta'}]+r^2[F_{\alpha\beta},
F_{\alpha'\beta'}] \cr &=& i{x_\alpha\over r^2}(2x_\beta
F_{\alpha'\beta'}+x_{\alpha'} F_{\beta\beta'}+x_{\beta'}
F_{\alpha'\beta})-i{x_\beta\over r^2} (2x_\alpha
F_{\alpha'\beta'}+x_{\alpha'} F_{\alpha\beta'}+x_{\beta'}
F_{\alpha'\alpha})\cr & &+r^2[F_{\alpha\beta}, F_{\alpha'\beta'}]
\cr &=& r^2[F_{\alpha\beta}, F_{\alpha'\beta'}]-{i\over
r^2}\left(-x_\alpha x_{\alpha'} F_{\beta\beta'}-x_\alpha x_{\beta'}
F_{\alpha'\beta}+x_\beta x_{\alpha'} F_{\alpha\beta'}+x_\beta
x_{\beta'} F_{\alpha'\alpha}\right)\cr & = &
i\delta_{\alpha\alpha'}F_{\beta\beta'}+i\delta_{\beta\beta'}F_{\alpha\alpha'}
-i\delta_{\alpha\beta'}F_{\beta\alpha'}-i\delta_{\beta\alpha'}F_{\alpha\beta'}.\nonumber
\end{eqnarray}
\end{proof}
By using Lemma \ref{lemma2} and the definition of $J_{\alpha\beta}$,
one can easily check that $J_{\alpha\beta}$'s satisfy the standard
commutation relation of $\frk{so}(D)$ Lie algebra. Then Lemma
\ref{lemma2} may be paraphrased as follows: under the commutation
action of $J_{\alpha\beta}$'s, $r$ and $1\over r$ transform as
$\frk{so}(D)$ scalars, $x_\alpha$'s and $\pi_\alpha$'s transform as
$\frk{so}(D)$ vectors, and $F_{\alpha\beta}$'s transform as a
$\frk{so}(D)$ bi-vectors. It is then clear that  $T$, $\Gamma_{D+1}$
and $\Gamma_{-1}$ transform as $\frk{so}(D)$ scalars; $\vec A$,
$\vec M$ and $\vec \Gamma$ transform as $\frk{so}(D)$ vectors. This
completes the proof for Eq. (\ref{cmtr}) in the case when it
involves $J_{\alpha\beta}$.

By using identities $x_\alpha{\mathcal A}_\alpha=0$ and $x_\alpha
F_{\alpha\beta}=0$, one can check that $[-\vec r\cdot \nabla, \vec
r]=-\vec r$,  $[-\vec r\cdot \nabla, r]=-r$, $[-\vec r\cdot \nabla,
{1\over r}]={1\over r}$, $[-\vec r\cdot \nabla, \vec \pi]=\vec \pi$.
That is, $-\vec r\cdot \nabla$ is the dimension operator in physics.
It is then clear that
\begin{eqnarray}
[\Gamma_{-1},T] = -i\Gamma_{D+1},\hskip 10pt[\Gamma_{D+1},T] =-
i\Gamma_{-1},\hskip 10pt [\vec \Gamma, T]=\vec 0. \end{eqnarray}
Consequently,
\begin{eqnarray}
[M_{\alpha}, T] & = & [i[\Gamma_\alpha, \Gamma_{-1}], T] =
i[\Gamma_\alpha,[\Gamma_{-1},
T]]+i[[\Gamma_\alpha,T],\Gamma_{-1}]\cr & = &  [\Gamma_\alpha,
\Gamma_{D+1}]=-iA_{\alpha},\cr [A_{\alpha}, T] & = &
[i[\Gamma_\alpha, \Gamma_{D+1}], T] = i[\Gamma_\alpha,[\Gamma_{D+1},
T]]+i[[\Gamma_\alpha,T],\Gamma_{D+1}]\cr & = &  [\Gamma_\alpha,
\Gamma_{-1}]=-iM_{\alpha}.
\end{eqnarray}
This completes the proof of Eq. (\ref{cmtr}) in the case when it
involves $T$.

\vskip 10pt
The remaining verifications are divided into four cases.

\underline{Case 1}.
\begin{eqnarray}
[\Gamma_\alpha, \Gamma_\beta] = -iJ_{\alpha\beta},\hskip 10pt
[\Gamma_\alpha, \Gamma_{D+1}]= -iA_\alpha, \cr [\Gamma_\alpha,
\Gamma_{-1}] = -iM_\alpha,\hskip 10pt [\Gamma_{D+1},\Gamma_{-1} ] =
-iT
\end{eqnarray} which are just the defining relations. So case 1 is done.

\underline{Case 2}.
\begin{eqnarray}
[M_\alpha, \Gamma_\beta]= -i\eta_{\alpha\beta}\Gamma_{-1},\hskip
10pt [A_\alpha, \Gamma_\beta]=
-i\eta_{\alpha\beta}\Gamma_{D+1}\nonumber
\end{eqnarray}or equivalently
\begin{eqnarray}
[Z_\alpha, \Gamma_\beta]= -i\eta_{\alpha\beta}X,\hskip 10pt
[W_\alpha, \Gamma_\beta]= -i\eta_{\alpha\beta}Y.
\end{eqnarray}
\begin{proof}
Since $[W_\alpha, \Gamma_\beta]=[x_\alpha, r\pi_\beta]=ir
\delta_{\alpha\beta}=-i\eta_{\alpha\beta}Y$, we just need to verify
the first identity. Note that $[Z_\alpha, r]=2i\Gamma_\alpha$ (see
case 3 below), so
\begin{eqnarray}
[Z_\alpha, \Gamma_\beta] & = & r[Z_\alpha,
\pi_\beta]+2i\Gamma_\alpha\pi_\beta\cr
 & = & r[x_\alpha\pi^2 - 2 \pi_\alpha(\vec r\cdot \vec
\pi)+2r^2F_{\alpha \beta}\pi_\beta-{c\over r^2}x_\alpha +i(D-3)
\pi_\alpha, \pi_\beta]+2ir\pi_\alpha\pi_\beta\cr & = &
r\left(i\delta_{\alpha\beta}\pi^2 -2ix_\alpha
F_{\gamma\beta}\pi_\gamma\right)+r(2i F_{\alpha\beta}(\vec r\cdot
\vec \pi)-2i\pi_\alpha\pi_\beta)+r(-2ir^2F_{\alpha
\gamma}F_{\gamma\beta}+[2r^2F_{\alpha \gamma},
\pi_\beta]\pi_\gamma)\cr & &+c r[\pi_\beta,{x_\alpha\over r^2}]
+(D-3)rF_{\alpha\beta}+2ir\pi_\alpha\pi_\beta\cr
 & = &
i\delta_{\alpha\beta}r\pi^2 -2irx_\alpha
F_{\gamma\beta}\pi_\gamma+2i rF_{\alpha\beta}(\vec r\cdot \vec
\pi)+r(-2ir^2F_{\alpha \gamma}F_{\gamma\beta}+4ix_\beta F_{\alpha
\gamma}\pi_\gamma-2r^2 [\pi_\beta, F_{\alpha \gamma}]\pi_\gamma)\cr
& &+c r[\pi_\beta,{x_\alpha\over r^2}] +(D-3)rF_{\alpha\beta}\cr
 & = &
i\delta_{\alpha\beta}r\pi^2 -2ir^3F_{\alpha
\gamma}F_{\gamma\beta}+2ir(2x_\beta F_{\alpha \gamma}+x_\alpha
F_{\beta \gamma}+x_\gamma F_{\alpha\beta}+r^2 [\nabla_\beta,
F_{\alpha \gamma}])\pi_\gamma\cr & &+c r[\pi_\beta,{x_\alpha\over
r^2}] +(D-3)rF_{\alpha\beta}\cr & = & i\delta_{\alpha\beta}r\pi^2
+2ir^3F_{ \gamma\alpha}F_{\gamma\beta}-ic
r\left(\delta_{\alpha\beta}-2{x_\alpha x_\beta\over
r^4}\right)+2(n-1)rF_{\alpha\beta}\cr &=&
i\delta_{\alpha\beta}(r\pi^2+{c\over
r^2})=-i\eta_{\alpha\beta}X.\nonumber
\end{eqnarray}
\end{proof}

\underline{Case 3}.
\begin{eqnarray}
[M_\alpha, \Gamma_{-1}]= i\Gamma_\alpha,\hskip 10pt [M_\alpha,
\Gamma_{D+1}] = 0\cr [A_\alpha, \Gamma_{-1}] = 0,\hskip 10pt
[A_\alpha, \Gamma_{D+1}] = -i\Gamma_\alpha.\nonumber
\end{eqnarray} or equivalently,
\begin{eqnarray}
[\vec W, Y]=[\vec Z, X]=0,\hskip 10pt [\vec W, X]=[\vec Z, Y]=2i\vec
\Gamma.
\end{eqnarray}
\begin{proof} It is clear that $[\vec W, Y]=0$. Now $[W_\alpha,
X]=r[x_\alpha, \pi^2]=ir\{\pi_\beta,
\delta_{\alpha\beta}\}=2i\Gamma_\alpha$. Next, using the identity
$F_{\alpha\beta}x_\beta=0$, we have
\begin{eqnarray}
[Z_\alpha, Y]&=&[x_\alpha\pi^2 - 2 \pi_\alpha(\vec r\cdot \vec
\pi)+2r^2F_{\alpha \beta}\pi_\beta-{c\over r^2}x_\alpha +i(D-3)
\pi_\alpha, r]\cr &=&x_\alpha[\pi^2, r] - 2 [\pi_\alpha(\vec r\cdot
\vec \pi), r]-2irF_{\alpha \beta}x_\beta +(D-3) {x_\alpha\over
r}\cr&=&-ix_\alpha\{\pi_\beta, {x_\beta\over r}\} - 2
\pi_\alpha[(\vec r\cdot \vec \pi), r] - 2 [\pi_\alpha, r](\vec
r\cdot \vec \pi) +(D-3) {x_\alpha\over r}\cr&=&-(D-1){x_\alpha\over
r}-2i{x_\alpha\over r}\vec r\cdot\vec \pi+2i \pi_\alpha r + 2i
{x_\alpha\over r}(\vec r\cdot \vec \pi) +(D-3) {x_\alpha\over
r}\cr&=&-2{x_\alpha\over r}+2i \pi_\alpha r
=2ir\pi_\alpha=2i\Gamma_\alpha.\nonumber
\end{eqnarray}
Finally,
\begin{eqnarray}
[{1\over r}, Z_\alpha] & = &  [{1\over r}, x_\alpha\pi^2 - 2
\pi_\alpha(\vec r\cdot \vec \pi)+2r^2F_{\alpha
\beta}\pi_\beta-{c\over r^2}x_\alpha +i(D-3) \pi_\alpha]\cr & = &
x_\alpha[{1\over r},\pi^2] - 2 [{1\over r},\pi_\alpha(\vec r\cdot
\vec \pi)]+2r^2F_{\alpha \beta}[{1\over r},\pi_\beta]+i(D-3)[{1\over
r},\pi_\alpha]\cr & = & -ix_\alpha\{{x_\beta\over r^3},\pi_\beta\} -
2 \pi_\alpha[{1\over r},(\vec r\cdot \vec \pi)]- 2 [{1\over
r},\pi_\alpha](\vec r\cdot \vec \pi)-2ir^2F_{\alpha
\beta}{x_\beta\over r^3}+(D-3){x_\alpha\over r^3}\cr & =
&-ix_\alpha[\pi_\beta,{x_\beta\over r^3}] -2i{x_\alpha\over r^3}\vec
r\cdot \vec \pi + 2i \pi_\alpha {1\over r}+ 2i {x_\alpha\over
r^3}(\vec r\cdot \vec \pi)+(D-3){x_\alpha\over r^3}\cr &=& 2i
\pi_\alpha {1\over r};\nonumber
\end{eqnarray}
\begin{eqnarray}
[r\pi^2, Z_\alpha] & = &  [r, Z_\alpha]\pi^2+r[\pi^2, Z_\alpha]\cr &
= &  -2i\Gamma_\alpha\pi^2+r[\pi^2, Z_\alpha]\cr
&=&-2i\Gamma_\alpha\pi^2+r[\pi^2,  x_\alpha\pi^2 - 2 \pi_\alpha(\vec
r\cdot \vec \pi)+2r^2F_{\alpha \beta}\pi_\beta-{c\over r^2}x_\alpha
+i(D-3) \pi_\alpha]\cr &=& -2i\Gamma_\alpha\pi^2+ r\left([\pi^2,
x_\alpha]\pi^2 - 2 [\pi^2, \pi_\alpha(\vec r\cdot \vec
\pi)]+2[\pi^2, r^2F_{\alpha \beta}\pi_\beta]\right)\cr
&&+r\left(-c[\pi^2, {x_\alpha\over r^2}] +i(D-3)[\pi^2,
\pi_\alpha]\right)\cr &=& -2i\Gamma_\alpha\pi^2+
r\left(-2i\pi_\alpha\pi^2 - 2 [\pi^2, \pi_\alpha](\vec r\cdot \vec
\pi)+4i\pi_\alpha\pi^2+2[\pi^2, r^2F_{\alpha
\beta}\pi_\beta]\right)\cr &&+r\left(-c[\pi^2, {x_\alpha\over r^2}]
+i(D-3)[\pi^2, \pi_\alpha]\right)\cr &=& r\left( - 2 [\pi^2,
\pi_\alpha](\vec r\cdot \vec \pi)+2[\pi^2, r^2F_{\alpha
\beta}]\pi_\beta+2r^2F_{\alpha \beta}[\pi^2, \pi_\beta]-c[\pi^2,
{x_\alpha\over r^2}] +i(D-3)[\pi^2, \pi_\alpha]\right)\cr
 &=& r\left( - 2 [\pi^2,
\pi_\alpha](\vec r\cdot \vec \pi)-2[\pi^2,
x_\alpha\pi_\beta-x_\beta\pi_\alpha]\pi_\beta\right)\cr
&&+r\left(2r^2F_{\alpha \beta}[\pi^2, \pi_\beta]-c[\pi^2,
{x_\alpha\over r^2}] +i(D-3)[\pi^2, \pi_\alpha]\right)\cr &=&
r\left( - 2 [\pi^2, \pi_\alpha](\vec r\cdot \vec \pi)-2[\pi^2,
x_\alpha\pi_\beta]\pi_\beta+2[\pi^2,x_\beta\pi_\alpha]\pi_\beta\right)\cr
&&+r\left(2r^2F_{\alpha \beta}[\pi^2, \pi_\beta]-c[\pi^2,
{x_\alpha\over r^2}] +i(D-3)[\pi^2, \pi_\alpha]\right)\cr&=& r\left(
- 2 [\pi^2, \pi_\alpha](\vec r\cdot \vec \pi)-2x_\alpha[\pi^2,
\pi_\beta]\pi_\beta+2x_\beta[\pi^2,\pi_\alpha]\pi_\beta+4F_{\alpha\beta}\pi_\beta\right)\cr
&&+r\left(2r^2F_{\alpha \beta}[\pi^2, \pi_\beta]-c[\pi^2,
{x_\alpha\over r^2}] +i(D-3)[\pi^2, \pi_\alpha]\right).\nonumber
\end{eqnarray}

To continue we note that $[\pi^2,
\pi_\alpha]=2iF_{\alpha\gamma}\pi_\gamma$, so
\begin{eqnarray}
[r\pi^2, Z_\alpha] &=& r\left( - 4i F_{\alpha\gamma}\pi_\gamma(\vec
r\cdot \vec \pi)-4ix_\alpha
F_{\beta\gamma}\pi_\gamma\pi_\beta+4ix_\beta
F_{\alpha\gamma}\pi_\gamma\pi_\beta+4F_{\alpha\beta}\pi_\beta\right)\cr
&&+r\left(4ir^2F_{\alpha \beta}F_{\beta\gamma}\pi_\gamma-c[\pi^2,
{x_\alpha\over r^2}] -2(D-3)F_{\alpha\gamma}\pi_\gamma]\right)\cr
&=& r\left( 2x_\alpha F_{\beta\gamma}F_{\beta\gamma}+4ir^2F_{\alpha
\beta}F_{\beta\gamma}\pi_\gamma-c[\pi^2, {x_\alpha\over r^2}]
-2(D-3)F_{\alpha\gamma}\pi_\gamma]\right)\cr &=& 4c_2{x_\alpha \over
r^3}-cr[\pi^2, {x_\alpha\over r^2}]+4ir\left(r^2F_{\alpha
\beta}F_{\beta\gamma}+i{D-3\over
2}F_{\alpha\gamma}\right)\pi_\gamma\cr &=& 4c_2{x_\alpha \over
r^3}+icr\{\pi_\beta, [\nabla_\beta,{x_\alpha\over
r^2}]\}-4ir{c_2\over n}\left({\delta_{\alpha\gamma}\over
r^2}-{x_\alpha x_\gamma\over r^4}\right)\pi_\gamma\cr &=&
4c_2{x_\alpha \over r^3}+icr\{\pi_\beta, {\delta_{\alpha\beta}\over
r^2}-2{x_\alpha x_\beta\over r^4}\}+4irc\left(-{1\over
r^2}\pi_\alpha+{x_\alpha \over r^4}\vec r\cdot \vec \pi\right)\cr
&=& 4c_2{x_\alpha \over r^3}+icr[\pi_\beta,
{\delta_{\alpha\beta}\over r^2}-2{x_\alpha x_\beta\over r^4}]-2ic
{1\over r}\pi_\alpha.\nonumber
\end{eqnarray}
Therefore,
\begin{eqnarray}
[X, Z_\alpha] &=& [r\pi^2+{c\over r}, Z_\alpha]=4c_2{x_\alpha \over
r^3}+icr[\pi_\beta, {\delta_{\alpha\beta}\over r^2}-2{x_\alpha
x_\beta\over r^4}]-2ic [{1\over r},\pi_\alpha]\cr &=& 4nc{x_\alpha
\over r^3}-2c(D-2){x_\alpha\over r^3}-2c{x_\alpha\over
r^3}=0.\nonumber
\end{eqnarray}
\end{proof}

\underline{Case 4}.
\begin{eqnarray}
[M_{\alpha},M_{\beta}] = -iJ_{\alpha\beta},\hskip 10pt
[A_{\alpha},M_{\beta}] = -i\eta_{\alpha\beta}T,\hskip 10pt
[A_{\alpha},A_{\beta}] = iJ_{\alpha\beta}\nonumber
\end{eqnarray}
or equivalently,
\begin{eqnarray}
[Z_{\alpha},Z_{\beta}] =[W_\alpha, W_\beta]= 0,\hskip 10pt
[Z_{\alpha},W_{\beta}]
=-2i\left(\eta_{\alpha\beta}T+J_{\alpha\beta}\right).
\end{eqnarray}
\begin{proof} It is clear that $[W_\alpha, W_\beta]= 0$ because
$W_\alpha=x_\alpha$. Next,
\begin{eqnarray}
[Z_{\alpha},W_{\beta}] & = & [x_\alpha \pi^2 - 2\pi_\alpha (\vec
r\cdot \vec \pi)+2r^2F_{\alpha  \gamma}\pi_\gamma-{c\over
r^2}x_\alpha +i(D-3)\pi_\alpha , x_\beta]\cr &=&x_\alpha [\pi^2,
x_\beta] - 2[\pi_\alpha (\vec r\cdot \vec \pi),
x_\beta]+2r^2F_{\alpha \gamma}[\pi_\gamma, x_\beta]
+i(D-3)[\pi_\alpha , x_\beta] \cr &=&-2ix_\alpha \pi_\beta -
2\pi_\alpha [(\vec r\cdot \vec \pi), x_\beta]- 2[\pi_\alpha,
x_\beta] (\vec r\cdot \vec \pi)-2ir^2F_{\alpha \beta}
+(D-3)\delta_{\alpha\beta} \cr &=&-2ix_\alpha \pi_\beta
+2i\pi_\alpha x_\beta+2i\delta_{\alpha\beta} (\vec r\cdot \vec
\pi)-2ir^2F_{\alpha\beta} +(D-3)\delta_{\alpha\beta}
 \cr &=&-2i\left(x_\alpha
\pi_\beta-x_\beta\pi_\alpha+r^2
F_{\alpha\beta}\right)+2i\delta_{\alpha\beta} \left (\vec r\cdot
\vec \pi-i{D-1\over 2}\right)\cr & = &
-2i\left(\eta_{\alpha\beta}T+J_{\alpha\beta}\right).\nonumber
\end{eqnarray}Finally, using results from case 2 and case 3, we have
\begin{eqnarray}
-i[Z_\alpha, Z_\beta]& = & [[\Gamma_\alpha, X], Z_\beta]=
[\Gamma_\alpha X-X\Gamma_\alpha, Z_\beta]\cr & = & [\Gamma_\alpha X,
Z_\beta]-[X\Gamma_\alpha, Z_\beta]=[\Gamma_\alpha,
Z_\beta]X-X[\Gamma_\alpha, Z_\beta]\cr &=& [[\Gamma_\alpha,
Z_\beta], X]=[i\eta_{\alpha\beta}X, X]=0.\nonumber
\end{eqnarray}
\end{proof}

End of the proof of part 1) of Theorem \ref{keyr}.

\subsection{Proof of part 2)}
We just need to verify equality
\begin{eqnarray} \sum_{1\le A\le D+1}\{J_{AB}, J_{AC}\}-\sum_{-1\le A\le 0}\{J_{AB}, J_{AC}\}=2a\eta_{BC}
\end{eqnarray}
under the condition that $B\le C$, to be more specific, we need to
verify the following identities:
\begin{eqnarray}\left\{\begin{array}{rcl}
 \sum_{1\le \alpha\le D}\{J_{\alpha\beta}, J_{\alpha\gamma}\}+\{A_\beta, A_\gamma\}-\{M_\beta,M_\gamma\}-\{\Gamma_\beta, \Gamma_\gamma\} & = &
 2a\eta_{\beta\gamma},\cr
\sum_{1\le \alpha\le D}\{J_{\alpha\beta},
A_{\alpha}\}-\{M_\beta,T\}-\{\Gamma_\beta, \Gamma_{D+1}\} & = &
 0,\cr
\sum_{1\le \alpha\le D} A_{\alpha}^2-T^2- \Gamma_{D+1}^2 & = &
 -a,\cr
  \sum_{1\le \alpha\le D}\{J_{\alpha\beta}, M_{\alpha}\}-\{A_\beta, T\}-\{\Gamma_\beta, \Gamma_{-1}\} & = &
 0,\cr
 \sum_{1\le \alpha\le D}\{A_{\alpha}, M_{\alpha}\}-\{\Gamma_{D+1}, \Gamma_{-1}\} & = &
 0,\cr
 \sum_{1\le \alpha\le D} M_{\alpha}^2+T^2- \Gamma_{-1}^2 & = &
 a,\cr
   \sum_{1\le \alpha\le D}\{J_{\alpha\beta}, \Gamma_{\alpha}\}-\{A_\beta, \Gamma_{D+1}\}+\{M_\beta, \Gamma_{-1}\} & = &
 0,\cr
 \sum_{1\le \alpha\le D}\{A_{\alpha}, \Gamma_{\alpha}\}+\{T, \Gamma_{-1}\} & = &
 0,\cr
 \sum_{1\le \alpha\le D}\{M_{\alpha}, \Gamma_{\alpha}\}+\{\Gamma_{D+1}, T\} & = &
 0,\cr
 \sum_{1\le \alpha\le D} \Gamma_{\alpha}^2+\Gamma_{D+1}^2- \Gamma_{-1}^2 & = &
 a.\end{array}\right.
\end{eqnarray}
The checking is then divided into six cases.

\underline{Case 1}.\begin{eqnarray}
 \sum_{1\le \alpha\le D} \Gamma_{\alpha}^2+\Gamma_{D+1}^2- \Gamma_{-1}^2 & = &
 a.
\end{eqnarray}
\begin{proof}
\begin{eqnarray}
 \sum_{1\le \alpha\le D} \Gamma_{\alpha}^2+\Gamma_{D+1}^2- \Gamma_{-1}^2 & =
 &r\pi_\alpha r\pi_\alpha-{1\over 2}(XY+YX)\cr
 &=& r^2\pi^2-i\vec r\cdot \vec \pi-{1\over
 2}(r\pi^2r+c+r^2\pi^2+c)\cr
 &=& -i\vec r\cdot \vec \pi-{1\over
 2}r[\pi^2,r]-c=-i\vec r\cdot \vec \pi+{i\over
 2}r\{\pi_\mu,{x_\mu\over r}\}-c\cr
 &=& {i\over
 2}r[\pi_\mu,{x_\mu\over r}]-c={D-1\over 2}-c =a.\nonumber
\end{eqnarray} \end{proof}

\underline{Case 2}.\begin{eqnarray}
 \sum_{1\le \alpha\le D}\{A_{\alpha}, \Gamma_{\alpha}\}+\{T,
 \Gamma_{-1}\}=0,\hskip 10pt
 \sum_{1\le \alpha\le D}\{M_{\alpha}, \Gamma_{\alpha}\}+\{\Gamma_{D+1},
 T\}=0;\nonumber
\end{eqnarray}
Or equivalently
\begin{eqnarray}
 \sum_{1\le \alpha\le D}\{Z_{\alpha}, \Gamma_{\alpha}\}+\{
 X, T\}=0,\hskip 10pt
 \sum_{1\le \alpha\le D}\{W_{\alpha}, \Gamma_{\alpha}\}-\{Y,
 T\}=0.
\end{eqnarray}
\begin{proof}
We check the 2nd identity first:
\begin{eqnarray}
 \sum_{1\le \alpha\le D}\{W_{\alpha}, \Gamma_{\alpha}\}-\{Y,
 T\}& = & \{x_\alpha, r\pi_\alpha\}-\{r, \vec r\cdot \vec \pi-{D-1\over 2}i\}\cr
 &=& 2r\vec r\cdot \vec \pi+r[\pi_\alpha,x_\alpha]-2r  \vec r\cdot \vec
 \pi-[\vec r\cdot \vec \pi,  r]+i(D-1)r\cr
 &= & 0.\nonumber
\end{eqnarray}
Then we check the 1st identity:
\begin{eqnarray}
 \sum_{1\le \alpha\le D}\{Z_{\alpha}, \Gamma_{\alpha}\}+\{
 X, T\} & = & 2\Gamma_\alpha Z_\alpha +2TX+[Z_\alpha, \Gamma_\alpha]+[X, T]\cr
 & = & 2(r\pi_\alpha x_\alpha\pi^2
- 2 r\pi^2(\vec r\cdot \vec \pi)+2r\pi_\alpha r^2F_{\alpha
\beta}\pi_\beta - r\pi_\alpha{c\over r^2}x_\alpha +i(D-3) r\pi^2)\cr
&& +2(\vec r\cdot \vec \pi-{D-1\over 2}i)(r\pi^2+{c\over
r})-i\eta_{\alpha\alpha}X-iX\cr
 & = & 2\left(r(\vec r\cdot \pi) \pi^2
- 2 r\pi^2(\vec r\cdot \vec \pi)+2 r^3F_{\alpha
\beta}\pi_\alpha\pi_\beta - cr\pi_\alpha{x_\alpha\over r^2} -3i
r\pi^2\right)\cr && +2\vec r\cdot \vec \pi(r\pi^2+{c\over r})\cr
 & = & 2\left(2ir\pi^2
- r\pi^2(\vec r\cdot \vec \pi)-i r^3F_{\alpha \beta}F_{\alpha\beta}
- {c\over r}\vec r\cdot \vec \pi- cr[\pi_\alpha,{x_\alpha\over r^2}]
-3i r\pi^2\right)\cr && +2\vec r\cdot \vec \pi(r\pi^2+{c\over r})\cr
& = & 2\left( [\vec r\cdot \vec \pi, r\pi^2]-i r^3F_{\alpha
\beta}F_{\alpha\beta} + [\vec r\cdot \vec \pi, {c\over r}]-
cr[\pi_\alpha,{x_\alpha\over r^2}] -i r\pi^2\right)\cr & = & 2\left(
-i r^3F_{\alpha \beta}F_{\alpha\beta} + i{c\over r}+ic {D-2\over r}
\right) \cr & = & 2\left( -2i {c_2\over r} + ic {D-1\over r}
\right)=0.\nonumber
\end{eqnarray}
\end{proof}

\underline{Case 3}.
\begin{eqnarray}
  && \sum_{1\le \alpha\le D}\{J_{\alpha\beta}, \Gamma_{\alpha}\}-\{A_\beta, \Gamma_{D+1}\}+\{M_\beta,
  \Gamma_{-1}\}=0.
\end{eqnarray}
\begin{proof}
\begin{eqnarray}
  && \sum_{1\le \alpha\le D}\{J_{\alpha\beta}, \Gamma_{\alpha}\}-\{A_\beta, \Gamma_{D+1}\}+\{M_\beta, \Gamma_{-1}\}\cr & =
   &2J_{\alpha\beta}\Gamma_{\alpha}+[\Gamma_\alpha, J_{\alpha\beta}]
+{1\over 2}\left(\{X, W_\beta\}+\{Y, Z_\beta\}\right)\cr &=&
2(x_\alpha\pi_\beta r \pi_\alpha-x_\beta\pi_\alpha r
\pi_\alpha+r^2F_{\alpha\beta}r\pi_\alpha)-i(D-1)\Gamma_\beta\cr & &
+X W_\beta+YZ_\beta +2i\Gamma_\beta \cr
 &=&
2(\pi_\beta r \vec r\cdot \vec \pi+[x_\alpha, \pi_\beta
r]\pi_\alpha-x_\beta[\pi_\alpha, r] \pi_\alpha-x_\beta
r\pi^2+r^3F_{\alpha\beta}\pi_\alpha)-i(D-3)\Gamma_\beta\cr & &
+(r\pi^2+{c\over r})x_\beta+rx_\beta \pi^2  - 2r\pi_\beta (\vec
r\cdot \vec \pi)+2r^3F_{\beta  \gamma}\pi_\gamma -{c\over r}x_\beta
+i(D-3)r\pi_\beta \cr &=& 2(\pi_\beta r \vec r\cdot \vec
\pi+ir\pi_\beta+i{x_\beta\over r}\vec r\cdot \pi)-x_\beta r\pi^2\cr
& & +r\pi^2 x_\beta - 2r\pi_\beta(\vec r\cdot \vec \pi)  \cr &=&
2(r\pi_\beta  \vec r\cdot \vec \pi+ir\pi_\beta) +[r\pi^2, x_\beta] -
2r\pi_\beta(\vec r\cdot \vec \pi)\cr &=& 2([r\pi_\beta,  \vec r\cdot
\vec \pi]+ir\pi_\beta) -2ir\pi_\beta=0.\nonumber
\end{eqnarray}
\end{proof}

\underline{Case 4}.
\begin{eqnarray}
\sum_{1\le \alpha\le D} A_{\alpha}^2-T^2- \Gamma_{D+1}^2 & = &
 -a,\cr
 \sum_{1\le \alpha\le D} M_{\alpha}^2+T^2- \Gamma_{-1}^2 & = &
 a,\cr \sum_{1\le \alpha\le D}\{A_{\alpha}, M_{\alpha}\}-\{\Gamma_{D+1},
\Gamma_{-1}\}& = &
 0;\nonumber
\end{eqnarray}
or equivalently
\begin{eqnarray}
\sum_{1\le \alpha\le D}Z_\alpha^2 =  X^2,\hskip 10pt  \sum_{1\le
\alpha\le D} \{Z_\alpha, W_\alpha\}+4T^2- \{X, Y\}= 4a.
\end{eqnarray}Here we have used the fact that $\sum_{1\le \alpha\le
D}W_\alpha^2=Y^2$.
\begin{proof}
To check the 1st identity, we note that $Z_\alpha=i[\Gamma_\alpha,
X]$ and $[Z_\alpha, X]=0$, so
\begin{eqnarray}
Z_\alpha^2={i\over 2}[\{Z_\alpha, \Gamma_\alpha\}, X].\nonumber
\end{eqnarray}
Then
\begin{eqnarray}
\sum_{1\le \alpha\le D}Z_\alpha^2 & = & {i\over 2}\sum_{1\le
\alpha\le D}[\{Z_\alpha, \Gamma_\alpha\}, X]\cr & = & -{i\over
2}[\{X, T\}, X]\hskip 20pt\mbox{use results from case 2}\cr &=&
-{i\over 2}[T, X^2]=X^2.\nonumber
\end{eqnarray}

To check the 2nd identity, we note that $Z_\alpha=i[\Gamma_\alpha,
X]$ and $[W_\alpha, X]=2i\Gamma_\alpha$, so
\begin{eqnarray}
\{Z_\alpha, W_\alpha\}=i[\{W_\alpha, \Gamma_\alpha\},
X]+4\Gamma_\alpha^2.\nonumber
\end{eqnarray}
Then
\begin{eqnarray}
\sum_{1\le \alpha\le D}\{Z_\alpha, W_\alpha\} & = & i[\{Y, T\},
X]+4\sum_{1\le \alpha\le D}\Gamma_\alpha^2\hskip 20pt \mbox{use
results from case 2}\cr & = & i\left(\{Y,[T,X]\}+\{[Y,
X],T\}\right)+4\sum_{1\le \alpha\le D}\Gamma_\alpha^2\cr &=&
i\left(\{Y,iX\}+\{2iT,T\}\right)+4\sum_{1\le \alpha\le
D}\Gamma_\alpha^2\cr &=&-\{X, Y\}-4T^2+4\sum_{1\le \alpha\le
D}\Gamma_\alpha^2.\nonumber
\end{eqnarray} So
\begin{eqnarray}
\sum_{1\le \alpha\le D}\{Z_\alpha, W_\alpha\} +4T^2-\{X, Y\}
&=&4\sum_{1\le \alpha\le D}\Gamma_\alpha^2-2\{X, Y\}\cr &=&4a \hskip
20pt \mbox{use results from case 1}.\nonumber
\end{eqnarray}\end{proof}

\underline{Case 5}.
\begin{eqnarray}
\sum_{1\le \alpha\le D}\{J_{\alpha\beta},
A_{\alpha}\}-\{M_\beta,T\}-\{\Gamma_\beta, \Gamma_{D+1}\} & = &
 0,\cr
  \sum_{1\le \alpha\le D}\{J_{\alpha\beta}, M_{\alpha}\}-\{A_\beta, T\}-\{\Gamma_\beta, \Gamma_{-1}\}
  & = &
 0;\nonumber
\end{eqnarray}
or equivalently
\begin{eqnarray}\left\{\begin{array}{rcl}
\sum_{1\le \alpha\le D}\{J_{\alpha\beta},
Z_{\alpha}\}-\{Z_\beta,T\}-\{\Gamma_\beta,X\} & = &
 0,\\
  \sum_{1\le \alpha\le D}\{J_{\alpha\beta}, W_{\alpha}\}+\{W_\beta, T\}-\{\Gamma_\beta, Y\}
  & = & 0.\end{array}\right.
\end{eqnarray}
\begin{proof}
We check the 2nd identity first:
\begin{eqnarray}
 && \sum_{1\le \alpha\le D}\{J_{\alpha\beta}, W_{\alpha}\}+\{W_\beta, T\}-\{\Gamma_\beta, Y\}\cr
   & = &
 2x_\alpha J_{\alpha\beta}+[J_{\alpha\beta}, x_\alpha]+2x_\beta\vec r\cdot\vec \pi+[\vec r\cdot\vec \pi,x_\beta]-i(D-1)x_\beta-2r^2\pi_\beta-r[\pi_\beta, r]\cr
 & = & 2r^2\pi_\beta-2x_\beta \vec r\cdot \vec \pi+[J_{\alpha\beta}, x_\alpha]+2x_\beta\vec r\cdot\vec \pi+[\vec r\cdot\vec \pi,x_\beta]
 -i(D-1)x_\beta-2r^2\pi_\beta-r[\pi_\beta, r]=0.\nonumber
 \end{eqnarray}
To check the 1st identity, we note that $Z_\alpha=i[\Gamma_\alpha,
X]$ and $[J_{\alpha\beta}, X]=0$, so
\begin{eqnarray} \sum_{1\le
\alpha\le D}\{J_{\alpha\beta}, Z_{\alpha}\} & =
&i[\{J_{\alpha\beta}, \Gamma_\alpha\}, X] \cr &=&-i[XW_\beta+Z_\beta
Y, X]\hskip 20pt\mbox{Use results from case 3}\cr &=&-iX[W_\beta,
X]-i[Z_\beta , X]Y-iZ_\beta[Y, X]\cr &=&2X\Gamma_\beta+2Z_\beta T
\hskip 20pt\mbox{Use results from commutation relations}\cr
&=&\{X,\Gamma_\beta\}+\{Z_\beta, T\}+[X,\Gamma_\beta]+[Z_\beta,
T]\cr &=&\{X,\Gamma_\beta\}+\{Z_\beta,
T\}+iZ_\beta-iZ_\beta=\{X,\Gamma_\beta\}+\{Z_\beta, T\}.\nonumber
\end{eqnarray} So the 1st identity is checked.
\end{proof}

\underline{Case 6}.
\begin{eqnarray}
 \sum_{1\le \alpha\le D}\{J_{\alpha\beta}, J_{\alpha\gamma}\}+\{A_\beta, A_\gamma\}-\{M_\beta,M_\gamma\}-\{\Gamma_\beta, \Gamma_\gamma\} & = &
 2a\eta_{\beta\gamma}.
 \end{eqnarray}
 \begin{proof}
\begin{eqnarray}
 \sum_{1\le \alpha\le D}\{J_{\alpha\beta}, J_{\alpha\gamma}\}&=& i\sum_{1\le \alpha\le D}\{J_{\alpha\beta}, [\Gamma_\alpha,
 \Gamma_\gamma]\}\cr
 &=& i\sum_{1\le \alpha\le D}\left([\{J_{\alpha\beta},\Gamma_\alpha\}, \Gamma_\gamma]-\{[J_{\alpha\beta},  \Gamma_\gamma], \Gamma_\alpha
\}\right)\cr &=& -i[X W_\beta+Z_\beta
Y,\Gamma_\gamma]-\{\Gamma_\alpha\delta_{\beta\gamma}-\Gamma_\beta\delta_{\alpha\gamma},\Gamma_\alpha\}\cr
&=& -iX [W_\beta,\Gamma_\gamma] -iZ_\beta [Y,\Gamma_\gamma] -i[X
,\Gamma_\gamma]W_\beta-i[Z_\beta ,\Gamma_\gamma]Y\hskip 10pt
\mbox{use results from case 3}\cr &&-2\delta_{\beta\gamma}\sum_{1\le
\alpha\le D}\Gamma_\alpha^2+\{\Gamma_\beta,\Gamma_\gamma\}\cr &=&
-\eta_{\beta\gamma} YX +Z_\beta W_\gamma + Z_\gamma
W_\beta-\eta_{\beta\gamma}YX\cr &&+2\eta_{\beta\gamma}\sum_{1\le
\alpha\le D}\Gamma_\alpha^2+\{\Gamma_\beta,\Gamma_\gamma\}.\nonumber
 \end{eqnarray}
So
\begin{eqnarray}
 \sum_{1\le \alpha\le D}\{J_{\alpha\beta}, J_{\alpha\gamma}\}-\{\Gamma_\beta,\Gamma_\gamma\} &=&\eta_{\beta\gamma}\left(2\sum_{1\le
\alpha\le D}\Gamma_\alpha^2-2YX\right) +Z_\beta W_\gamma + Z_\gamma
W_\beta\cr & = & \eta_{\beta\gamma} \left(2\sum_{1\le \alpha\le
D}\Gamma_\alpha^2-2YX\right) +{1\over 2}\left(\{Z_\beta, W_\gamma\}
+ \{Z_\gamma, W_\beta\}\right)\cr && -{1\over 2}\left([Z_\beta,
W_\gamma]+[Z_\gamma, W_\beta]\right)\cr &=&\eta_{\beta\gamma}
\left(2\sum_{1\le \alpha\le D}\Gamma_\alpha^2-2YX+2iT\right)
+{1\over 2}\left(\{Z_\beta, W_\gamma\} + \{Z_\gamma,
W_\beta\}\right) \cr &=&\eta_{\beta\gamma} \left(2\sum_{1\le
\alpha\le D}\Gamma_\alpha^2-\{X, Y\}\right) -\{A_\beta,
A_\gamma\}+\{M_\beta, M_\gamma\}.\nonumber
 \end{eqnarray}
So the identity is true because in case 1 we have verified that
\begin{eqnarray}
 2\sum_{1\le \alpha\le
D}\Gamma_\alpha^2-\{X, Y\}=2a.\nonumber
 \end{eqnarray}
 \end{proof}

End of the proof of part 2) of Theorem \ref{keyr}.

\section{Representation theoretical aspects --- the preliminary part}\label{Rep}
The main objective in the rest of this paper is to show that the
algebraic direct sum $\mathcal H$ of the energy eigenspaces of a
generalized MICZ-Kepler problem in dimension $(2n+1)$ is a unitary
highest weight $(\frk{g}, K)$-module where $\frk{g}=\frk{so}(2n+4)$
and $K=\mr{Spin}(2)\times_{{\bb Z}_2}\mr{Spin}(2n+2)$. Along the
way, we prove Theorem \ref{main}.

We can label the generators of $\frk{g}_0$ (the Lie algebra of
$\mr{Spin}(2,2n+2)$) as follows:
$$
M_{AB}=-M_{BA} \quad \mbox{for $A, B = -1, 0, 1,  \ldots, 2n+2$ }
$$ where in the $(2n+4)$-dimensional defining representation, the
matrix elements of $M_{AB}$ are given by
$$
[M_{AB}]_{JK}=-i(\eta_{AJ}\eta_{BK}-\eta_{BJ}\eta_{AK})
$$ with the indefinite metric tensor $\eta$ being ${\mr
{diag}}\{++-\cdots-\}$ relative to the following order: $-1$, $0$,
$1$, \ldots, $2n+2$ for the indices.

One can easily show that
\begin{eqnarray}\label{cmtrm} [M_{AB},
M_{A'B'}]=i(\eta_{AA'}M_{BB'}+\eta_{BB'}M_{AA'}-\eta_{AB'}M_{BA'}-\eta_{BA'}M_{AB'}).
\end{eqnarray}

In view of the sign difference between the right hand sides of Eqs.
(\ref{cmtr}) and (\ref{cmtrm}), we define the representation
$(\tilde\pi,C^\infty({\mathcal S}^{2\mu}))$ of $\frk{g}$ as follows:
for $\psi\in C^\infty({\mathcal S}^{2\mu})$, \begin{eqnarray}\fbox{$
\tilde\pi(M_{AB})(\psi)= -\hat J_{A B}\psi$}
\end{eqnarray}
where, by definition, $\hat J_{AB}:={1\over {\sqrt r}}J_{AB}\sqrt
r$.

However, what is really relevant for us is just a subspace of
$C^\infty({\mathcal S}^{2\mu})$, i.e., $\mathcal H$. Actually, the
story is bit more involved: what is really invariant under
$\tilde\pi$ is not $\mathcal H$, but a twisted version of $\mathcal
H$ which is denoted by $\tilde {\mathcal H}$ later; and there is a
twist linear equivalence
$$\tau:\;\mathcal H \to \tilde{\mathcal H}$$ which preserves the $L^2$-norm,
such that, viewing $\tau$ as an equivalence of representations, we
get representation $(\pi, \mathcal H)$. Because of this intricacy,
we shall devote the next two subsections to some preparations.

\subsection{Review of the (bound) energy eigenspaces} The bound eigen-states (i.e., \emph{$L^2$ eigen-sections} of the Hamiltonian) of
the generalized MICZ-Kepler problems have been analyzed in section
5.1 of Ref. \cite{meng05} by using the classical analytic method
with the help of the representation theory for compact Lie groups.
Recall that the (bound) energy spectrum is
\begin{eqnarray}
E_I=-{1\over 2(I+n+|\mu|)^2}
\end{eqnarray} where $I=0, 1, 2, \cdots$.

Denote by ${\mathcal S}^{2\mu}|_{\mr{S}^{2n}}$ the restriction
bundle of ${\mathcal S}^{2\mu}$ to the unit sphere $\mr{S}^{2n}$. As
a hermitian bundle with a hermitian connection, ${\mathcal
S}^{2\mu}|_{\mr{S}^{2n}}$ is just the vector bundle
$$\mr{Spin}(2n+1)\times_{\mr{Spin}(2n)} {\bf s}^{2\mu}\to
\mr{S}^{2n}$$ with the natural $\mr{Spin}(2n+1)$-invariant
connection. Note that, as a hermitian bundle with a hermitian
connection, ${\mathcal S}^{2\mu}$ is the pullback of ${\mathcal
S}^{2\mu}|_{\mr{S}^{2n}}$ under the natural projection ${\bb
R}_*^{2n+1}\to \mr{S}^{2n}$. Let $L^2({\mathcal S}^{2\mu})$,
$L^2({\mathcal S}^{2\mu}|_{\mr{S}^{2n}})$ be the $L^2$-sections of
${\mathcal S}^{2\mu}$ and ${\mathcal S}^{2\mu}|_{\mr{S}^{2n}}$
respectively. It is clear that $\mr{Spin}(2n+1)$ acts on both
$L^2({\mathcal S}^{2\mu})$ and $L^2({\mathcal
S}^{2\mu}|_{\mr{S}^{2n}})$ unitarily. In fact, as a representation
of $\mr{Spin}(2n+1)$, $L^2({\mathcal S}^{2\mu}|_{\mr{S}^{2n}} )$ is
the induced representation of ${\bf s}^{2\mu}$ from $\mr{Spin}(2n)$
to $\mr{Spin}(2n+1)$; therefore, by the Frobenius reciprocity plus a
branching rule\footnote{See, for example, Theorem 2 of \S 129 of
Ref. \cite{DZ73}. } for $(\mr{Spin}(2n+1), \mr{Spin}(2n))$, one has
\begin{eqnarray}\label{brch0}
L^2({\mathcal S}^{2\mu}|_{\mr{S}^{2n}})=\hat\bigoplus_{l\ge 0}{\ms
R}_l
\end{eqnarray} where ${\ms R}_l$ is the irreducible representation of
$\mr{Spin}(2n+1)$ with highest weight $(l+|\mu|, |\mu|, \cdots,
|\mu|)$. Observe that, if we use $\widetilde X$ to denote the
horizontal lift of vector field $X$ on ${\bb R}^{2n+1}_*$, then the
vector field $[\widetilde{r\partial_\alpha},
\widetilde{r\partial_\beta}]$ can be shown to be just the right
invariant vector field on ${\bb R}_+\times \mr{Spin}(2n+1)$ whose
value at $(r, e)$ (where $e$ is the group identity element) is
$(0,-i\gamma_{\alpha\beta})$, i.e., $(0,-{1\over 4}[e_\alpha,
e_\beta])$. Consequently, the infinitesimal action of
$\mr{Spin}(2n+1)$ on $C^\infty({\mathcal S}^{2\mu})$ is just the
restriction of $\tilde\pi$ to $\mr{span}_{\bb
R}\{M_{\alpha\beta}\mid 1\le\alpha<\beta\le
2n+1\}=\frk{so}_0(2n+1)$. It is then clear that
$\tilde\pi(M_{\alpha\beta})$'s act only on the angular part of the
wave sections --- a consequence which can also be deduced from the
fact that $\hat J_{\alpha\beta}$'s commute with the multiplication
by a smooth function of $r$.

Let $\{Y_{l\bf m}(\Omega)\}_{{\bf m}\in {\mathcal I}(l)}$ be an
orthornormal (say Gelfand-Zeltin) basis for ${\ms R}_l$, and
$$\fbox{$l_\mu=l+|\mu|+n-1$}.$$ Then, an orthornormal basis for the
energy eigenspace $\ms H_I$ with energy $E_I$ is
\begin{eqnarray}
\{\psi_{kl\bf m}:=R_{kl_\mu}(r)Y_{l\bf m}(\Omega)\; |\; k+l=I+1,
k\ge 1, l\ge 0, {\bf m}\in {\mathcal I}(l)\}
\end{eqnarray} where $R_{kl_\mu}\in L^2({\bb R}_+, r^{2n}\,dr)$ is a square integrable (with respect to measure $r^{2n}\,dr$) solution of
the radial Schr\"{o}dinger equation:
\begin{eqnarray}\label{rSchEq}
\left(-{1\over 2r^{2n}}\partial_r
r^{2n}\partial_r+{l_\mu(l_\mu+1)-n(n-1)\over 2r^2}-{1\over
r}\right)R_{kl_\mu}=E_{k-1+l}R_{kl_\mu}.
\end{eqnarray}
Note that $R_{kl_\mu}$ is of the form
$$r^{-n}y_{kl_\mu}(r)\exp\left(-{r\over k+l_\mu}\right)$$
with $y_{kl_\mu}(r)$ satisfying Eq.
\begin{eqnarray}
\left( {d^2\over dr^2} -{2\over k+l_\mu} {d\over dr}+\left[{2\over
r}-{l_\mu(l_\mu+1)\over
r^2}\right]\right)y_{kl_\mu}(r)=0.\end{eqnarray}In term of the
generalized Laguerre polynomials,
$$
y_{kl_\mu}(r)=c(k,l) r^{l_\mu+1}L^{2l_\mu+1}_{k-1}\left({2\over
k+l_\mu}r\right)
$$
where $c(k,l)$ is a constant, which can be uniquely determined by
requiring $c(k,l)>0$ and $\int_0^\infty |R_{kl_\mu}(r)|^2
r^{2n}dr=1$.

We are now ready to state the following remark.
\begin{rmk}\label{rmk1}
1) $\ms H_I$ is the space of square integrable solutions of Eq. $
H\psi = E_I\psi$.

2) As representation of $\frk{so}(2n+1)$,
\begin{eqnarray}
{\ms H}_I=\bigoplus_{l=0}^I D_l
\end{eqnarray} where $D_l:=\mr{span}\{\psi_{(I-l+1)l\bf m}\mid
{\bf m}\in {\mathcal I}(l)\}$ is the highest weight module with
highest weight $(l+|\mu|, |\mu|, \cdots, |\mu|)$.

3) $\{{\ms H}_I \mid I= 0, 1, 2, \ldots\}$ is the complete set of
(bound) energy eigenspaces.

\end{rmk}

For the completeness of this review, we state part of Theorem 1 from
Ref. \cite{meng05} below:
\begin{Th}  For the $(2n+1)$-dimensional generalized
MICZ-Kepler problem with magnetic charge $\mu$, the following
statements are true:

1) The negative energy spectrum is
$$
E_I=-{1/2\over (I+n+|\mu|)^2}
$$ where $I=0$, $1$, $2$, \ldots;

2) The Hilbert space $\ms H(\mu)$ of negative-energy states admits a
linear $\mr{Spin}(2n+2)$-action under which there is a decomposition
$$
{\ms H}(\mu)=\hat\bigoplus _{I=0}^\infty\,{\ms H}_I
$$ where ${\ms H}_I$ is the irreducible $\mr{Spin}(2n+2)$-module
with highest weight $(I+|\mu|,|\mu|, \cdots, |\mu|, \mu)$;

3) The linear action in part 2) extends the manifest linear action
of $\mr{Spin}(2n+1)$, and ${\ms H}_I$ in part 2) is the energy
eigenspace with eigenvalue $E_I$ in part 1).
\end{Th}

It was shown in Ref. \cite{meng05} that the bound eigen-states are
precisely the ones with negative energy eigenvalues. We would like
to remark that, in dimension five, a similar result obtained with a
similar method has already appeared in Ref. \cite{Levay00}.

\subsection{Twisting} As we said before, because of the technical
intricacy, we need to introduce the notion of twisting. Let us start
with the listing of some important spaces used later:
\begin{itemize}
\item $\ms H_I$ --- the $I$-th bound energy eigenspace;
\item $\mathcal H$ --- the algebraic direct sum of all bound energy eigenspaces;
\item $\ms H$ or $\ms H(\mu)$ --- the completion of $\mathcal H$ under the standard $L^2$-norm;
\item ${\mathcal H}_{l\bf m}$ --- the subspace of $\mathcal H$ spanned by $\{\psi_{kl\bf m}\,|\, k\ge
1,\; \mbox{$l$, $\bf m$ fixed}\}$;
\item ${\ms H}_{l\bf m}$ --- the completion of ${\mathcal H}_{l\bf
m}$ under the standard $L^2$-norm.
\end{itemize}
Note that these spaces are all endowed with the unique hermitian
inner product which yields the standard $L^2$-norm, i.e.,
\begin{eqnarray}
\fbox{$\langle\psi,\phi\rangle:=\displaystyle\int_{{\bb
R}^{D}_*}(\psi, \phi)\, d^{D}x$}
\end{eqnarray} where $(\psi, \phi)$ is the point-wise hermitian inner product
and $d^Dx$ is the Lebesgue measure.

It is clear from the previous section that
\begin{eqnarray}
{\ms B}:=\{\psi_{kl\bf m}\; |\;k\ge 1, l\ge 0, {\bf m}\in {\mathcal
I}(l)\}
\end{eqnarray} is an orthonormal basis for both ${\mathcal H}$ and $\ms H$.

To study the action of $\hat J_{AB}$'s, we need to ``twist" $\ms B$,
$\ms H_I$, ${\mathcal H}_{l\bf m}$, ${\ms H}_{l\bf m}$, $\mathcal H$
and $\ms H$ to get $\tilde{\ms B}$, $\tilde {\ms H}_I$,
$\tilde{\mathcal H}_{l\bf m}$, $\tilde{\ms H}_{l\bf m}$, $\tilde
{\mathcal H}$ and $\tilde {\ms H}$ respectively. It suffices to
twist the elements of ${\ms B}$. Let $\tau$: $\ms B\to\tilde{\ms B}$
be defined as follows:
\begin{eqnarray}\label{psitilde}
\tau(\psi_{kl\bf m})(r,
\Omega)&:=&(k+l_\mu)\,e^{-i\theta_{k+l_\mu}\hat T}\left({1\over
\sqrt r}\psi_{kl\bf m}(r, \Omega)\right)\cr  &=&
(k+l_\mu)^{n+1}{1\over \sqrt r}\psi_{kl\bf m}((k+l_\mu)r, \Omega)\cr
& \varpropto & r^{l +|\mu|-{1\over
2}}\,L^{2l_\mu+1}_{k-1}(2r)\,e^{-r}\,Y_{l\bf m}(\Omega)
\end{eqnarray}
where $\hat T={1\over \sqrt r} T \sqrt r$, and $\theta_I=-\ln I$ for
any positive number $I$. For simplicity, we write $\tau(\psi_{kl\bf
m})$ as $\tilde\psi_{kl\bf m}$. One can check that
$$
\int_{{\bb R}^{D}_*}(\tilde \psi_{kl\bf m}, \tilde \psi_{kl\bf m})\,
d^{D}x=\int_{{\bb R}^{D}_*}(\psi_{kl\bf m}, \psi_{kl\bf m})\,
d^{D}x=1.
$$
By using Eq. (\ref{psitilde}) and the orthogonality identities for
the generalized Laguerre polynomials, one can see that
$\tilde\psi_{kl\bf m}$ is orthogonal to $\tilde\psi_{k'l\bf m}$ when
$k\neq k'$.

It is now clear how to twist all the relevant spaces listed in the
beginning of this subsection. For example,
\begin{eqnarray}
{\tilde {\ms H}}_I:=\left\{e^{-i\theta_{I_\mu+1}\hat T}\left({1\over
\sqrt r}\psi\right)\,|\, \psi \in \ms H_I\right\}.
\end{eqnarray}
Since $\ms H_I$ is spanned by $\{\psi_{kl\bf m}\,|\, k+l=I+1, k\ge
1, l\ge 0, {\bf m}\in {\mathcal I}(l)\}$, it follows that
$\tilde{\ms H}_I$ is spanned by
$$\{\tilde \psi_{kl\bf m}\,|\, k+l=I+1, k\ge 1, l\ge 0, {\bf m}\in
{\mathcal I}(l)\}.$$

We shall call $\tilde{\ms H}(\mu)$ the twisted Hilbert space of the
bound states for the $(2n+1)$-dimensional generalized MICZ-Kepler
problem with magnetic charge $\mu$. Remark that the twisting
map\footnote{It has a \underline{basis-free} description: for
$\psi\in \ms H_I$, $\tau(\psi)(r, \Omega)=(I_\mu+1)^{n+1}{1\over
\sqrt r}\psi((I_\mu+1)r, \Omega)$. }
\begin{eqnarray}\tau:\; {\ms
H}(\mu)\to \tilde{\ms H}(\mu)
\end{eqnarray}
is the unique linear isometry which sends $\psi_{kl\bf m}$ to
$\tilde \psi_{kl\bf m}$; moreover, $\tau$ maps all relevant
subspaces of ${\ms H}(\mu)$ isomorphically onto the corresponding
relevant twisted subspaces. Note that $\hat J_{\alpha\beta}={1\over
\sqrt r} J_{\alpha\beta} \sqrt r= J_{\alpha\beta}$ obviously acts on
$\tilde {\ms H}_I$ as hermitian operator, so
$\frk{r}:=\mr{span}\{M_{\alpha\beta}\, |\, 1\le \alpha< \beta\le
2n+1\}=\frk{so}(2n+1)$  acts unitarily on $\tilde {\ms H}_I$ via
$\tilde \pi$.

Recall that for non negative integer $I$, we use $I_\mu$ to denote
$I+n+|\mu|-1$.
\begin{Prop}\label{Prmk2}
1) $\tilde\psi_{kl\bf m}$ is an eigenvector of $\hat \Gamma_{-1}$
with eigenvalue $k+l_\mu$.

2) $\tilde{\ms H_I}$ is the space of square integrable solutions of
Eq. $\hat\Gamma_{-1}\psi = (I_\mu+1)\psi$.

3) $\hat\Gamma_{-1}$ is a self-adjoint operator on $\tilde{\ms
H}(\mu)$ and $\tilde{\ms H_I}$ is the eigenspace of
$\hat\Gamma_{-1}$ with eigenvalue $I_\mu+1$.

4) As representation of $\frk{r}$,
\begin{eqnarray}
\tilde {\ms H}_I=\bigoplus_{l=0}^I \tilde D_l
\end{eqnarray} where $\tilde D_l:=\mr{span}\{\tilde \psi_{(I-l+1)l\bf m}\mid
{\bf m}\in {\mathcal I}(l)\}$ is the highest weight module with
highest weight $(l+|\mu|, |\mu|, \cdots, |\mu|)$.

5) $\tilde{\ms H}(\mu)= L^2({\mathcal S}^{2\mu})$.

\end{Prop}
\begin{proof}
1) The proof is based on the ideas from Ref. \cite{Barut71}. Since
\begin{eqnarray}\label{eigeneq1}
H\psi_{kl\bf m}=E_{k+l-1}\psi_{kl\bf m},
\end{eqnarray}
we have $ \sqrt r(H-E_{k+l-1})\psi_{kl\bf m}=0$ which can be
rewritten as
$$
({1\over 2}\hat X-1-E_{k+l-1}\hat Y)({1\over \sqrt r}\psi_{kl\bf
m})=0
$$ where $X$ and $Y$ are given by Eq. (\ref{def1}). In terms
of $\hat \Gamma_{-1}$ and $\hat \Gamma_{D+1}$, we can recast the
above equation as
$$
\left(({1\over 2}-E_{k+l-1})\hat \Gamma_{-1}+({1\over
2}+E_{k+l-1})\hat \Gamma_{D+1}-1\right)({1\over \sqrt r}\psi_{kl\bf
m})=0
$$
Plugging $\psi_{kl\bf m}={1\over k+l_\mu}\sqrt r
e^{i\theta_{k+l_\mu}\hat T}\left(\tilde \psi_{kl\bf m}\right)$ into
the above equation and using identities
\begin{eqnarray}
\left\{ \begin{array}{rcl}e^{-i\theta \hat
T}\,\hat\Gamma_{-1}\,e^{i\theta \hat T} & = &
\cosh\theta\,\hat\Gamma_{-1}+\sinh\theta\,\hat\Gamma_{D+1}\\
\\
e^{-i\theta \hat T}\,\hat\Gamma_{D+1}\,e^{i\theta \hat T} & = &
\sinh\theta\,\hat\Gamma_{-1}+
\cosh\theta\,\hat\Gamma_{D+1},\end{array}\right.
\end{eqnarray}
we arrive at the following equation:
\begin{eqnarray}\label{eigeneq2}
\fbox{$\hat \Gamma_{-1}\tilde\psi_{kl\bf
m}=(k+l_\mu)\tilde\psi_{kl\bf m}$}\,.
\end{eqnarray}

2) Note that the Barut-Bornzin process going from Eq.
(\ref{eigeneq1}) to Eq. (\ref{eigeneq2}) is completely reversible.
Therefore, part 2) is just a consequence of part 1) of Remark
\ref{rmk1}.

3) Note that $\hat \Gamma_{-1}$ is defined on the dense linear
subspace $\tilde{\mathcal H}$ of $\tilde{\ms H} (\mu)$. It is easy
to check that $$\langle\tilde\psi_{k'l'\bf m'}, \hat
\Gamma_{-1}\tilde\psi_{kl\bf m}\rangle=\langle\hat
\Gamma_{-1}\tilde\psi_{k'l'\bf m'}, \tilde\psi_{kl\bf m}\rangle$$
for any $\tilde\psi_{kl\bf m}$ and $\tilde\psi_{k'l'\bf m'}$.
Therefore, $\hat \Gamma_{-1}$ (To be precise, it should be its
closure) is a self-adjoint operator on $\tilde{\ms H} (\mu)$. In
view of part 2), $\tilde{\ms H}_I$ is the eigenspace of
$\hat\Gamma_{-1}$ with eigenvalue $I_\mu+1$.

4) This part is clear due to part 2) of Remark \ref{rmk1}.

5) Recall that $\tilde\psi_{kl\bf m}(r,\Omega)=\tilde
R_{kl_\mu}(r)\,Y_{l\bf m}(\Omega)$ where $\tilde
R_{kl_\mu}(r)\propto r^{l +|\mu|-{1\over
2}}\,L^{2l_\mu+1}_{k-1}(2r)\,e^{-r}$.  By the well-known property
for the generalized Laguerre polynomials, for any $l\ge 0$,
$\{\tilde R_{kl_\mu}\}_{k=1}^\infty$ form an orthonormal basis for
$L^2(\bb R_+, r^{2n}\,dr)$.

By virtue of Theorem II. 10 of Ref. \cite{Reed&Simon} and Eq.
(\ref{brch0}),
\begin{eqnarray}
L^2({\mathcal S}^{2\mu}) & = & L^2(\bb R_+, r^{2n}\,dr)\otimes
L^2({\mathcal S}^{2\mu}\mid_{\mr{S}^{2n}})\cr & = & \hat
\bigoplus_{l=0}^\infty \left(L^2(\bb R_+, r^{2n}\,dr)\otimes {\ms
R}_l\right).\nonumber
\end{eqnarray} Therefore, $\tilde {\ms B}$ is an orthonormal basis
for $L^2({\mathcal S}^{2\mu})$, consequently $\tilde{\ms H}(\mu)=
L^2({\mathcal S}^{2\mu})$.

\end{proof}
We end this subsection with
\begin{rmk}\label{rmk2}
$\tilde{\ms H}_I$ is the eigenspace of $\tilde\pi(H_0)$ with
eigenvalue $-(I_\mu+1)$. Here $\tilde\pi(H_0)=-\hat \Gamma_{-1}$ is
viewed as an endomorphism of $\tilde {\mathcal H}$.
\end{rmk}

\section{Representation theoretical aspects --- the final part}
We start with some notations:
\begin{itemize}
\item $G=\mr{Spin}(2, 2n+2)$ --- the double
cover of $\mr{SO}_0(2, 2n+2)$ characterized by the homomorphism
$\pi_1(\mr{SO}_0(2, 2n+2))=\bb Z\oplus \bb Z_2\to \bb Z_2$ sending
$(a, b)$ to $\bar a+b$;
\item $\frk{g}_0$ --- the Lie algebra of $\mr{Spin}(2, 2n+2)$;
\item $\frk{g}$ --- the complexfication of $\frk{g}_0$, so $\frk{g}=\frk{so}(2n+4)$;
\item $H_0$ --- defined to be $M_{-1, 0}$;
\item $H_j$ --- defined to be $-M_{2j-1, 2j}$ for $1\le j\le n+1$;
\item $K:=\mr{Spin}(2)\times_{\bb Z_2} \mr{Spin}(2n+2)$ --- a maximal compact subgroup of $\mr{Spin}(2, 2n+2)$;
\item $\frk{k}_0$ --- the Lie algebra of $K$;
\item $\frk{k}$ --- the complexfication of $\frk{k}_0$, so $\frk{k}=\frk{so}(2)\oplus\frk{so}(2n+2)$;
\item $\frk{r}$ --- the subalgebra of $\frk{g}$ generated by
$\{M_{AB}\,|\, 1\le A < B \le 2n+1\}$, so $\frk{r}_0:=\frk{g}_0\cap
\frk{r}=\frk{so}_0(2n+1)$;
\item $\frk{s}$ --- the subalgebra of $\frk{g}$ generated by  $\{M_{AB}\,|\, 1\le A < B
\le 2n+2\}$, so $\frk{s}_0:=\frk{g}_0\cap \frk{s}=\frk{so}_0(2n+2)$;
\item $\frk{sl}(2)$ --- the subalgebra of $\frk{g}$ generated by  $M_{-1, D+1}$, $M_{0, D+1}$ and $M_{-1,0}$, so $\frk{sl}_0(2):=\frk{g}_0\cap
\frk{sl}(2)=\frk{so}_0(2,1)$;
\item $U(\frk{sl}(2))$ --- the universal enveloping algebra of
$\frk{sl}(2)$.
\end{itemize}

\subsection{$\tilde{\mathcal H}$ is a unitary highest weight
Harish-Chandra module} The goal of this subsection is to show that
$(\tilde\pi, \tilde{\mathcal H})$ is a unitary highest weight
$(\frk{g}, K)$-module.

\begin{Prop}\label{prop1}
1) Each $\tilde\pi(M_{AB})$ maps $\tilde{\mathcal H}$ into
$\tilde{\mathcal H}$, so $(\tilde \pi, \tilde{\mathcal H})$ is a
representation of $\frk{g}$.

2) Each $\tilde\pi(M_{AB})$ is a hermitian operator on
$\tilde{\mathcal H}$, so $(\tilde \pi, \tilde{\mathcal H})$ is a
unitary representation of $\frk{g}$.

3) $(\tilde \pi|_{\frk{sl}(2)}, \tilde{\mathcal H}_{l\bf m})$ is the
discrete series representation of $\frk{so}(2,1)$ with highest
weight $-l_\mu-1$.

\end{Prop}
\begin{proof} 1) We follow the convention of Ref. \cite{georgi82} for describing the root space of $\frk{g}=\frk{so}(2n+4)$.
Take as a basis of the Cartan sub-algebra of $\frk{g}$ the following
elements:
$$
H_0=M_{-1, 0},\quad H_j=-M_{2j-1, 2j},\quad \mbox{$j=1, \cdots,
n+1$}.
$$
Let $\eta, \eta'=\pm 1$. We take the following root vectors:
$$
E_{\eta e^j+\eta' e^k}={1\over 2}\left(M_{2j-1, 2k-1}+i\eta M_{2j,
2k-1}+i\eta'M_{2j-1, 2k}-\eta\eta' M_{2j, 2k}\right)
$$ where $0\le j<k\le n+1$. This way we obtain a
Cartan basis for $\frk{g}$. Therefore, for $\psi_I\in \tilde{\ms
H}_I$, we have
\begin{eqnarray}\label{eigeneq}
\tilde \pi(H_0)(\tilde\pi(E_\alpha)(\psi_I))
 & = & (-I_\mu-1+\alpha_0)\tilde\pi(E_\alpha)(\psi_I)\cr & = &
 (-(I-\alpha_0)_\mu-1)\tilde\pi(E_\alpha)(\psi_I)
\end{eqnarray}
where $\alpha_0$ (which can be $0$, or $-1$ or  $1$) is the $0$-th
component of $\alpha$. It is not hard to see that
$\tilde\pi(E_\alpha)(\psi_I)$ is square integrable \footnote{The
convergence of the integral near infinity is clear because of the
exponential decay as $r\to \infty$. The convergence of the integral
near the origin of $\bb R^D_*$ in the general case is clear from the
counting of powers of $r$, and the remaining case when $n=1$, $\mu
=0$ and $\psi_I=\tilde \psi_{k00}$ needs a separate but equally easy
argument.}, so in view of part 2) of Proposition \ref{Prmk2}, Eq.
(\ref{eigeneq}) implies that $\tilde\pi(E_\alpha)(\psi_I)\in
{\tilde{\ms H}}_{I-\alpha_0}$. (Here ${\ms H}_{-1}=0$.) Therefore,
$\tilde\pi(E_\alpha)$ maps any $\tilde{\ms H}_I$, hence
$\tilde{\mathcal H}$, into $\tilde{\mathcal H}$. By a similar
argument, one can show that $\tilde\pi(H_i)$ maps $\tilde{\mathcal
H}$ into itself. Since $H$'s and $E$'s form a basis for $\frk{g}$,
this implies that $\tilde\pi(M_{AB})$ maps $\tilde{\mathcal H}$ into
itself.

2) It is equivalent to checking that each $\hat J_{AB}:={1\over
{\sqrt r}}J_{AB}\sqrt r$ is an hermitian operator on
$\tilde{\mathcal H}$. First of all, it is not hard to see that, when
${\mathcal O} =\pi_\alpha, r, {1\over r}, \sqrt r, {1\over \sqrt
r}$, we always have
\begin{eqnarray}\label{hermitian}\langle \psi_1, {\mathcal
O}\psi_2\rangle=\langle {\mathcal O}\psi_1,
\psi_2\rangle\end{eqnarray} for any $\psi_1$, $\psi_2$ in
$\tilde{\mathcal H}$. It is equally easy to see that Eq.
(\ref{hermitian}) is always true for any $\psi_1$, $\psi_2$ in
$\tilde{\mathcal H}$ when ${\mathcal O}$ is $\hat\Gamma_\alpha
=\sqrt r \pi_\alpha \sqrt r$, $\hat X =\sqrt r \pi^2 \sqrt r+{c\over
r}$, or $\hat Y=r$. It is then clear from definitions (\ref{def1})
and (\ref{def2}) that Eq. (\ref{hermitian}) is always true for any
$\psi_1$, $\psi_2$ in $\tilde{\mathcal H}$ when ${\mathcal O} =\hat
J_{AB}$.

3) Let us first show that $\tilde\pi(M_{-1, D+1})$, $\tilde\pi(M_{0,
D+1})$ and $\tilde\pi(M_{-1,0})$ map each $\tilde\psi_{kl\bf m}$
into $\tilde{\mathcal H}_{l\bf m}$, so they indeed map
$\tilde{\mathcal H}_{l\bf m}$ into $\tilde{\mathcal H}_{l\bf m}$.
This is obvious for $\tilde\pi(M_{-1,0})$ because
$\tilde\pi(M_{-1,0})(\tilde\psi_{kl\bf m})= -\hat
\Gamma_{-1}\tilde\psi_{kl\bf m}=-(k+l_\mu)\tilde\psi_{kl\bf m}$.
Next, we introduce
$$E_\pm={1\over \sqrt 2}(M_{-1, D+1}\pm iM_{0, D+1}),$$then one can
check from Eq. (\ref{cmtrm}) that $[M_{-1,0}, E_\pm]=\pm E_\pm$.
Therefore
$$
\tilde\pi(M_{-1,0})(\tilde\pi(E_\pm)( \tilde\psi_{kl\bf
m}))=(-k-l_\mu\pm 1)\tilde\pi( E_\pm)( \tilde\psi_{kl\bf m}),
$$ where $\tilde\pi( E_\pm)={1\over \sqrt 2}(\hat T\pm i\hat\Gamma_{D+1})$. It is not hard to see that $ \tilde\pi( E_\pm)(
\tilde\psi_{kl\bf m})$ is square integrable. In view of part 2) of
Proposition \ref{Prmk2}, we conclude that $\tilde \pi( E_\pm)(
\tilde\psi_{kl\bf m})$ must be proportional to $\tilde\psi_{(k\mp
1)l\bf m}$. (Here, by convention, $\tilde\psi_{0l\bf m}=0$.)
Therefore, operators $\tilde \pi( E_\pm)$ map $\tilde\psi_{kl\bf m}$
into $\tilde{\mathcal H}_{l\bf m}$. This proves that $(\tilde
\pi|_{\frk{sl}(2)}, \tilde{\mathcal H}_{l\bf m})$ is a
representation of $\frk{sl}(2)$.

In view of the fact that $\tilde\pi(M_{-1, 0})(\tilde \psi_{1l\bf
m})=-(l_\mu+1)\tilde \psi_{1l\bf m}\neq 0$, we conclude that
$U(\frk{sl}(2))\cdot \tilde \psi_{1l\bf m}$ is a \emph{nontrivial}
unitary highest weight representation of the non-compact real Lie
algebra $\frk{sl}_0(2)$, hence must be the discrete series
representation with highest weight $-(l_\mu+1)$. Since
$U(\frk{sl}(2))\cdot \tilde \psi_{1l\bf m}\subset \tilde {\mathcal
H}_{l\bf m}$, and $\dim ({\ms H}_I\,\cap\,U(\frk{sl}(2))\cdot \tilde
\psi_{1l\bf m})=\dim ({\ms H}_I\,\cap\, \tilde {\mathcal H}_{l\bf
m})$ for all $I\ge 0$, we conclude that $U(\frk{sl}(2))\cdot \tilde
\psi_{1l\bf m}= \tilde {\mathcal H}_{l\bf m}$. Therefore, $\tilde
{\mathcal H}_{l\bf m}$ is a unitary highest weight
$\frk{sl}(2)$-module with highest weight $-l_\mu-1$, which in fact
is a unitary highest weight $(\frk{sl}(2), \mr{Spin}(2))$-module.
Then $\tilde {\ms H}_{l\bf m}$ must be the discrete series
representation of $\mr {Spin}(2,1)$ with highest weight $-l_\mu-1$.

\end{proof}

To continue the discussion on representations, we prove the
following proposition.
\begin{Prop}\label{prop2}

1) $(\tilde\pi|_{\frk{s}}, \tilde{\ms H}_I)$ is an irreducible
unitary representation of $\frk{s}$, in fact, it is the highest
weight representation with highest weight $(I+|\mu|, |\mu|, \ldots,
|\mu|, \mu)$.

2) The unitary action of $\frk{k}_0$ on $\tilde{\mathcal H}$ can be
lifted to a unique unitary action of $K$ under which
\begin{eqnarray}
\tilde {\mathcal H} =\bigoplus_{l=0}^\infty \left(D(-l_\mu-1)\otimes
D^l\right)
\end{eqnarray} where $D^l$ is the irreducible module of $\mr{Spin}(2n+2)$
with highest weight $(l+|\mu|, |\mu|, \cdots, |\mu|, \mu)$ and
$D(-l_\mu-1)$ is the irreducible module of $\mr{Spin}(2)$ with
weight $-l_\mu-1$.

3) $\tilde{\mathcal H}$ is a unitary $(\frk{g}, K)$-module.

4) $(\tilde\pi, \tilde{\mathcal H})$ is irreducible; in fact, it is
the unitary highest weight module of $\frk{g}$ with highest weight
$$(-(n+|\mu|),|\mu|, \cdots, |\mu|, \mu).$$
\end{Prop}
\begin{proof}
1) Recall that $\frk{s}$ is the $\frk{so}(2n+2)$ Lie sub algebra of
$\frk{g}$ generated by $$ \{H_i, E_{\pm e^j\pm e^k}\,|\, 1\le i \le
n+1, 1\le j < k \le n+1\}, $$  and $\frk{s}_0:=\frk{s}\cap
\frk{g}_0$ is the compact real form of $\frk{s}$. Since $H_0$
commutes with any element in $\frk{s}$, in view of Remark
\ref{rmk2}, we conclude that each $\tilde{\ms H}_I$ is invariant
under $\tilde \pi(\frk{s})$, i.e., $(\tilde\pi|_{\frk{s}},
\tilde{\ms H}_I)$ is a representation of $\frk{s}$.

Inside $\frk{s}$ there is an $\frk{so}(2n+1)$ Lie sub algebra
$\frk{r}$. Note that $H_1$, \ldots, $H_n$ are the generators of a
Cartan subalgebra of $\frk{r}$, and $H_1$, \ldots, $H_{n+1}$ are the
generators of a Cartan subalgebra of $\frk{s}$. Recall from part 4)
of Proposition \ref{Prmk2},
\begin{eqnarray}\label{brch1}
(\tilde \pi|_{\frk{r}}, \tilde{\ms H}_I)=\bigoplus_{l=0}^I \tilde
D_l\end{eqnarray} where $\tilde D_l$ is the highest weight
$\frk{r}$-module with highest weight $(l+|\mu|, |\mu|, \cdots,
|\mu|)$.

By applying the branching rule\footnote{See, for example, Theorem 3
of \S 129 of Ref. \cite{DZ73}. } for $(\frk{s},\frk{r})$, one finds
that there are only two solutions to Eq. (\ref{brch1}):
$(\tilde\pi|_{\frk{s}},\tilde{\ms H}_I)$ is the highest weight
module of $\frk{s}$ with highest weight equal to either $(I+|\mu|,
|\mu|, \cdots, |\mu|, \mu)$ or $(I+|\mu|, |\mu|, \cdots, |\mu|,
-\mu)$. Let $\tilde\psi_{1I\bf I}\in \tilde{\ms H}_I$ be an
$\frk{s}$-highest weight vector, which is assumed to have unit norm.
Since $\tilde \pi(H_{n+1})=\hat A_D$, we have either $\hat
A_D\tilde\psi_{1I\bf I} =\mu \tilde\psi_{1I\bf I}$ or $\hat
A_D\tilde\psi_{1I\bf I} =-\mu\tilde\psi_{1I\bf I}$. To determine the
sign, we only need to  \underline{show that} $\langle
\tilde\psi_{1I\bf I},A_D\tilde\psi_{1I\bf I}\rangle=\mu$. Note that
$A_D=i[\Gamma_D, \Gamma_{D+1}]=i[\Gamma_D,
\Gamma_{-1}-r]=i[\Gamma_D, \Gamma_{-1}]-x_D$ and $\tilde\psi_{1I\bf
I}$ is an eigenvector of $\hat\Gamma_{-1}$, so
\begin{eqnarray} \langle \tilde\psi_{1I\bf
I},\hat A_D\tilde\psi_{1I\bf I}\rangle & = &  - \langle
\tilde\psi_{1I\bf I},x_D\tilde\psi_{1I\bf I}\rangle\cr
&=&-\int_{{\bb R}^{D}_*}x_D\left|\tilde \psi_{1I\bf I}(r,
\Omega)\right|^2\, d^{D}x.
\end{eqnarray}
One can show that\footnote{To be more specific, one needs to
generalize the work of Ref. \cite{WuYang76}. Since we are only
interested in a sign, we choose to skip the details here.}, up to a
multiplicative constant, $\tilde\psi_{1I\bf I}(r, \Omega)$ is equal
to
$$r^{I_\mu-n+{1\over
2}}\,e^{-r}\cdot (\sin\theta)^{-(n-1)}\,(1-\cos
\theta)^{I_\mu+\mu\over 2}\,(1+\cos \theta)^{I_\mu-\mu\over 2}\cdot
Z(\theta_1,\ldots, \theta_{D-3}, \phi).$$ Then
\begin{eqnarray} \langle \tilde\psi_{1I\bf
I},\hat A_D\tilde\psi_{1I\bf I}\rangle &=&-\int_{{\bb
R}^{D}_*}x_D\left|\tilde \psi_{1I\bf I}(r, \Omega)\right|^2\,
d^{D}x\cr
 &=& -{\displaystyle\int_0^\infty r^{2I_\mu+2}\,e^{-2r}\, dr\over \displaystyle\int_0^\infty
 r^{2I_\mu+1}\,e^{-2r}\,dr}\cdot
{\displaystyle\int_0^\pi \cos\theta\, (1-\cos
\theta)^{I_\mu+\mu}\,(1+\cos \theta)^{I_\mu-\mu}\sin\theta\,
d\theta\over \displaystyle\int_0^\pi (1-\cos
\theta)^{I_\mu+\mu}\,(1+\cos \theta)^{I_\mu-\mu}\sin\theta\,
d\theta} \cr
 &=& -{\Gamma(2I_\mu+3)\over 2\cdot \Gamma(2I_\mu+2)}\cdot {\displaystyle\int_{-1}^1
 x\,(1-x)^{I_\mu+\mu}\,(1+x)^{I_\mu-\mu}\,dx\over
\displaystyle\int_{-1}^1
 (1-x)^{I_\mu+\mu}\,(1+x)^{I_\mu-\mu}\,dx}\cr
 &=&  -(I_\mu+1)\cdot \left({\displaystyle\int_{-1}^1
 (1-x)^{I_\mu+\mu}\,(1+x)^{I_\mu+1-\mu}\,dx\over
\displaystyle\int_{-1}^1
 (1-x)^{I_\mu+\mu}\,(1+x)^{I_\mu-\mu}\,dx}-1\right)\cr
 &=&  -(I_\mu+1)\cdot \left(2\cdot {B(I_\mu+1+\mu,I_\mu+2-\mu)\over
 B(I_\mu+1+\mu,I_\mu+1-\mu)}-1\right)\cr
 &=&  -(I_\mu+1)\cdot \left(2\cdot {\Gamma(I_\mu+2-\mu)\Gamma(2I_\mu+2)\over
 \Gamma(I_\mu+1-\mu)\Gamma(2I_\mu+3)}-1\right)=\mu.\nonumber
\end{eqnarray}
Part 1) is done.

2) Since $\tilde{\ms H_l}$ is the space of square integrable
solutions of Eq. $ \hat\Gamma_{-1}\psi = (l_\mu+1)\psi$ and $\tilde
\pi(H_0)=-\hat\Gamma_{-1}$, as a $\frk{k}$-module, $\tilde{\ms
H_l}=D(-l_\mu-1)\otimes D^l$ where $D^l$ is the irreducible module
of $\mr{Spin}(2n+2)$ with highest weight $(l+|\mu|, |\mu|, \cdots,
|\mu|, \mu)$ and $D(-l_\mu-1)$ is the irreducible module of
$\mr{Spin}(2)$ with weight $-l_\mu-1$. Since $\mu$ is a half
integer, the irreducible unitary action of $\frk{k}_0$ on
$\tilde{\ms H_l}$ can be promoted to a unique irreducible unitary
action of $K$. Therefore, $\tilde{\mathcal H}$ is a unitary
$K$-module and has the following decomposition into isotypic
components of $K$:
$$
{\tilde{\mathcal H}}=\bigoplus_{l=0}^\infty \tilde{\ms
H}_l=\bigoplus_{l=0}^\infty \left(D(-l_\mu-1)\otimes D^l\right).
$$

3) From the definition, it is clear that the action of $K$ on
${\tilde{\mathcal H}}$ is compatible with that of $\frk{g}$ on
${\tilde{\mathcal H}}$, and its linearization agrees with the action
of $\frk{k}_0$. Part 2) says that ${\tilde{\mathcal H}}$ is
$K$-finite. Therefore, ${\tilde{\mathcal H}}$ is a unitary
$(\frk{g}, K)$-module.

4) Let $v\neq 0$ be a vector in $\tilde{\ms H}_0$ with
$\frk{g}$-weight $(-(n+|\mu|), |\mu|, \ldots, |\mu|, \mu)$. Since
this weight is the highest among all weights with a nontrivial
weight vector in $\tilde{\mathcal H}$, $V:=U(\frk{g})\cdot v\subset
\tilde{\mathcal H}$ is the unitary highest weight $\frk{g}$-module
with highest weight $(-(n+|\mu|), |\mu|, \ldots, |\mu|, \mu)$. Since
$\tilde{\ms H}_l$ is irreducible under $\frk{s}\subset \frk{g}$,
either $\tilde{\ms H}_l\subset V$ or $\tilde{\ms H}_l\cap V=0$, so
in particular $\tilde{\ms H}_0\subset V$. We \underline{claim} that
$\tilde{\ms H}_l\subset V$ for any $l\ge 0$, consequently
$V=\tilde{\mathcal H}$ and then part 4) is done. To prove the claim,
we note that $U(\frk{sl}(2))\cdot v$ must be the discrete series
representation of $\frk{sl}(2)$ with highest ($H_0$-) weight
$-(n+|\mu|)$ because it is a nontrivial unitary highest weight
representation of the non-compact Lie algebra $\frk{sl}_0(2)$. In
view of the fact that $\tilde{\ms H}_l$ is the eigenspace of
$\tilde\pi(H_0)$ with eigenvalue $-(l_\mu+1)$, $\tilde{\ms
H}_l\,\cap \,(U(\frk{sl}(2))\cdot
v)=\mr{span}\{\tilde\pi(E_-^l)(v)\}$ must be one-dimensional. Then
$\tilde{\ms H}_l\,\cap \,V\neq 0$ because $\dim (\tilde{\ms
H}_l\,\cap \,V)\ge \dim (\tilde{\ms H}_l\,\cap
\,(U(\frk{sl}(2))\cdot v))=1$.

\end{proof}
\subsection{Proof of Theorem \ref{main}}
Viewing the twisting map $\tau$ as an equivalence of
representations, we get a representation $\pi$ of $\frk{g}$
equivalent to $\tilde \pi$. Then the two propositions proved in the
previous subsection are true if we drop all ``tilde" there. Thus
${\mathcal H}$ is the unitary highest weight $(\frk{g}_0, K)$-module
with highest weight $(-(n+|\mu|), |\mu|, \ldots, |\mu|, \mu)$. By a
standard theorem of Harish-Chandra\footnote{See, for example,
Theorem 7 on page 71 of Ref. \cite{BK96}}, we know that ${\ms H}$ is
the unitary highest weight $G$-module with highest weight
$(-(n+|\mu|), |\mu|, \ldots, |\mu|, \mu)$ such that $(\pi, {\mathcal
H})$ is the underlying $(\frk{g}_0, K)$-module. One can check that
this highest weight module occurs at the first reduction point of
the Enright-Howe-Wallach classification diagram\footnote{Page 101,
Ref. \cite{EHW82}. In our case $z=A(\lambda_0)=n+1$. It is in Case
II when $\mu=0$. For $\mu\neq 0$, it is in Case I for $p=n+1$ or in
Case III depending on the sign of $\mu$. See pages 125 -126, Ref.
\cite{EHW82}. Note that, while there are two reduction points when
$\mu= 0$, there is only one reduction point when $\mu\neq 0$. }. So
part 1) is done. Part 2) of Theorem \ref{main} is just a consequence
of part 3) of Proposition \ref{prop1}, and part 3) of Theorem
\ref{main} is just a consequence of part 2) of Proposition
\ref{prop2}.

\appendix
\section{Geometrically transparent description}
The purpose of this appendix is to give a geometrically transparent
description of the unitary highest weight module of $\mr{Spin}(2,
2n+2)$ with highest weight $\left(-(n+|\mu|),|\mu|, \cdots, |\mu|,
\mu\right)$.

As usual, we assume $n\ge 1$ is an integer and let ${\mathcal
S}^{2\mu}$ be the pullback bundle under the natural retraction ${\bb
R}_*^{2n+1}\to \mr{S}^{2n}$ of the vector bundle
$\mr{Spin}(2n+1)\times_{\mr{Spin}(2n)} {\bf s}^{2\mu}\to
\mr{S}^{2n}$ with the natural $\mr{Spin}(2n+1)$-invariant
connection. Let $d^Dx$ be the Lebesgue measure on ${\bb R}^{2n+1}$.
As is standard in geometry, we use $ L^2({\mathcal S}^{2\mu})$ to
denote the Hilbert space of square integrable (with respect to
$d^Dx$) sections of ${\mathcal S}^{2\mu}$. We have shown that
$\tilde{\ms H}(\mu)=L^2({\mathcal S}^{2\mu})$, therefore,
$\left(\tilde \pi, L^2({\mathcal S}^{2\mu})\right)$ is the unitary
highest weight module of $\mr{Spin}(2, 2n+2)$ with highest weight
$\left(-(n+|\mu|),|\mu|, \cdots, |\mu|, \mu\right)$. To describe the
infinitesimal action of $\mr{Spin}(2, 2n+2)$ on $C^\infty({\mathcal
S}^{2\mu})$, it suffices to describe how $M_{\alpha,0}$, $M_{D+1,0}$
and $M_{-1, 0}$ act as differential operators. It is easy to see
that $M_{\alpha,0}$, $M_{D+1,0}$ and $M_{-1, 0}$ are equal to
$i\sqrt r\nabla_\alpha\sqrt r$, ${1\over 2}\left(\sqrt
r\Delta_\mu\sqrt r+r-{c\over r}\right)$ and ${1\over 2}\left(\sqrt
r\Delta_\mu\sqrt r-r-{c\over r}\right)$ respectively. Here
$\Delta_\mu$ is the Laplace operator twisted by ${\mathcal
S}^{2\mu}$. For example, for $\psi\in C^\infty({\mathcal
S}^{2\mu})$, we have
\begin{eqnarray}
(M_{\alpha,0}\cdot \psi)(r, \Omega) & = & i\sqrt r\nabla_\alpha
\left (\sqrt r\psi(r, \Omega)\right).
\end{eqnarray}

\end{document}